\documentclass[sigconf]{acmart}

\usepackage{subfig}
\usepackage[title]{appendix}
\usepackage{color}

\AtBeginDocument{%
  \providecommand\BibTeX{{%
    \normalfont B\kern-0.5em{\scshape i\kern-0.25em b}\kern-0.8em\TeX}}}

\setcopyright{acmcopyright}
\copyrightyear{2020}
\acmYear{2020}

\acmDOI{}
\acmConference[San Diego '20]{San Diego '20: 19th International Workshop on Data Mining in Bioinformatics}{August 24, 2020}{San Diego, CA}
\acmBooktitle{San Diego '20: 19th International Workshop on Data Mining in Bioinformatics,
  August 24, 2020, San Diego, CA}
\acmPrice{}
\acmISBN{}

\acmSubmissionID{7}

\begin{document}

\title{Scalable Bayesian Functional Connectivity Inference for Multi-Electrode Array Recordings}

\author{Yun Zhao, Richard Jiang, Zhenni Xu, Elmer Guzman, Paul K. Hansma, Linda Petzold}
\affiliation{%
  \institution{University of California, Santa Barbara}
  \streetaddress{}
  \city{Santa Barbara}
  \state{CA}
  \postcode{93106}
}
\email{yunzhao@cs.ucsb.edu}




\begin{abstract}

Multi-electrode arrays (MEAs) can record extracellular action potentials (also known as 'spikes') from hundreds or thousands of neurons simultaneously. Inference of a functional network from a spike train is a fundamental and formidable computational task in neuroscience. With the advancement of MEA technology, it has become increasingly crucial to develop statistical tools for analyzing multiple neuronal activity as a network. In this paper, we propose a scalable Bayesian framework for inference of functional networks from MEA data. Our framework makes use of the hierarchical structure of networks of neurons. We split the large scale recordings into smaller local networks for network inference, which not only eases the computational burden from Bayesian sampling but also provides useful insights on regional connections in organoids and brains. We speed up the expensive Bayesian sampling process by using parallel computing. Experiments on both synthetic datasets and large-scale real-world MEA recordings show the effectiveness and efficiency of the scalable Bayesian framework. Inference of networks from controlled experiments exposing neural cultures to cadmium presents distinguishable results and further confirms the utility of our framework. 

\end{abstract}




\maketitle

\section{Introduction}

Neuroscience deals with how networks of neurons are organized and how they function~\cite{neuroscience}.  Understanding connectivity between neurons and within the brain is a fundamental problem in neurobiology~\cite{functional}. Functional connectivity, defined as the statistical dependencies between different brain regions with similar patterns, is widely used in various neural tasks~\cite{individual,memory}. For instance, functional connectivity magnetic resonance imaging (MRI) is crucial for diagnosing and comprehending autism spectrum disorders~\cite{underconnected}. MEAs~\cite{meas} can record extracellular action potentials from hundreds or thousands of neurons and provide insights on neuronal connectivity~\cite{synaptic}. For hours or weeks, action potentials can be non-invasively monitored, when neurons are grown on planar MEAs~\cite{recording}. Further, there is a trend towards increasing the density of the arrays~\cite{highdensity} to better understand the neuron connectivities. 

MEA recordings provide researchers opportunities to understand neuron activities in many regions such as the brain, retina, and heart~\cite{liu2012use}. However, the analysis of this data is challenging, in part because of its high dimensionality. Summary statistics could be used to measure the connection weights between electrodes. These include, for example, Pearson correlation~\cite{garofalo2009evaluation}, cross correlogram (CCG), the maximal information coefficient (MIC)~\cite{MIC} as well as biophysically-inspired metrics~\cite{bio-inspired}. However, a data generative model is required to understand the underlying structure and to make full use of the domain expert knowledge~\cite{liu2017active}. Furthermore, these summary statistic methods provide different functional connectivity results for the same recording since they are all deterministic metrics, which present fixed connection weights between every two electrodes instead of a probabilistic estimation. 

Bayesian inference can address the requirements for the inference,  since it provides distributions for parameters using probabilistic models and observation data.  In contrast to deterministic optimization procedures that give point estimates of the unknown functional connectivity, computing a Bayesian posterior yields probability distributions for the neuronal network functional connectivity. Bayesian inference has been combined with the generalized linear model (GLM), with graph-based priors to infer the neuron connectivity pattern for analysis~\cite{Linderman}. However, there is a lack of scalable Bayesian techniques for inference of network structure, which is particularly acute for inference from high-density recordings. 
\begin{figure*}[ht]
  \includegraphics[width=\linewidth]{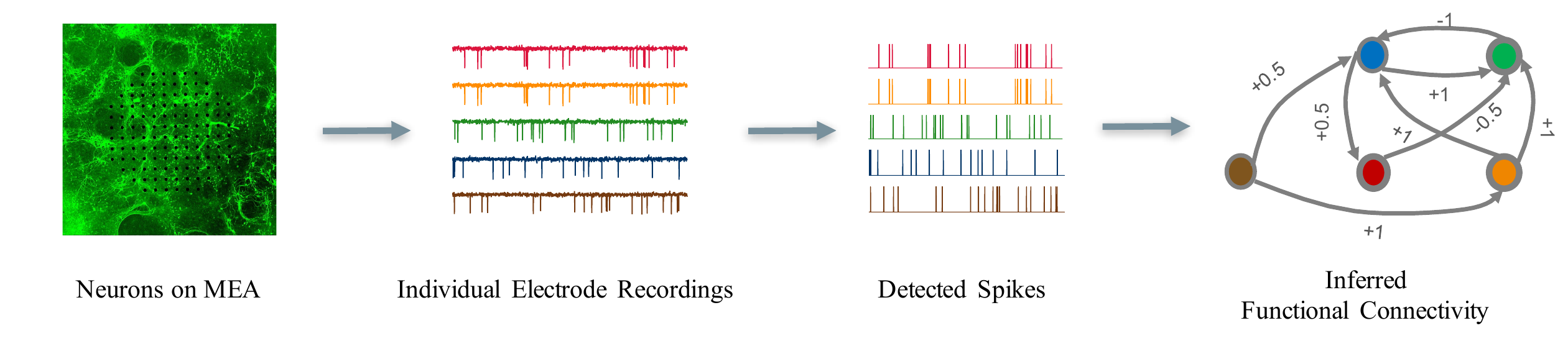}
  \caption{The workflow for Bayesian functional connectivity inference.  We consider only negative deflections that exceeded 6 times the standard deviation of the median noise level as spikes.}
  \label{framework}
\end{figure*}

A considerable challenge for Bayesian techniques is the rapid growth of computation time in accordance with the increasing scale of the network. In this paper, we propose a scalable framework of Bayesian inference, inspired by the hierarchical structure of networks of neurons.
Experiments on both synthetic datasets and large scale real-world MEA recordings show that our framework provides accurate and insightful results. Furthermore, we apply the proposed framework to a controlled cadmium dataset, and the results confirm its utility.

The key contributions of this paper include:

1) We propose a scalable functional connectivity inference framework shown in Fig.~\ref{framework} for MEA recording data. We speed up the expensive Bayesian sampling process through the use of parallel computing.

2) We infer the network by splitting the large scale recordings into smaller local networks. The splitting strategy decreases the average sampling time quadratically in accordance with the number of smaller local networks. We also provide a strategy for inferring the regional connectivity between local networks. This not only eases the computational burden from sampling but also provides useful insights on regional connections in organoids and brains.

3) Experiments on both synthetic dataset and large-scale real-world MEA recordings show the effectiveness and efficiency of the Bayesian framework.  Inference of network structure of Cadmium-exposed neuron cultures further demonstrates the usefulness of our framework. 

The remainder of this paper is organized as follows.  Section 2 describes the MEA data collection.  We delineate the probabilistic models in Section~\ref{model} and demonstrate the Bayesian inference details in Section~\ref{Bayesian}.  Section~\ref{split} describes the hierarchical setup.  Results for both synthetic and real data are provided in Sections~\ref{synthetic} and~\ref{real}, respectively. Related work is described in Section~\ref{related}. Section~\ref{discussion} is the Discussion.

\section{Data Collection}\label{Data}
\subsection{Cell Culture}
We prepared hippocampal neurons from postnatal day 0 (P0) mice with C57BL/6 genetic background using a previously described protocol~\cite{Tovar2018}. Cleaned and sterilized MEAs (120MEA100/30iR-ITO arrays; Multi Channel Systems) were incubated with poly-L-lysine (0.1 mg/ml) for at least one hour at 37 $^\circ$C, rinsed 3 times with sterile deionized water and allowed to air dry before cell plating. Glial cultures were maintained in separate T-75 flasks. 100, 000 - 125, 000 dissociated glial cells were used for the first plating of MEAs to obtain a confluent glial culture over the surface of the electrodes. Once glia were confluent, the hippocampi dissected from the brain followed by manual dissociation were plated at 250, 000 cells in the MEA chamber. Cultures were grown in a tissue culture incubator (37 $^\circ$C, 5\% $CO_{2}$) in a medium made with minimum essential medium + Earle’s salts (Thermo Scientific, catalog \# 11090081) with 2mM Glutamax (Thermo Scientific), 5\% heat-inactivated fetal bovine serum (Thermo Scientific), and 1 ml/l Mito+ serum extender (Corning) and supplemented with glucose to an added concentration of 21mM. To minimize the effects of evaporation, maintain cell culture sterility, and decrease degassing of the medium during recordings, the MEA chamber was covered by a gas permeable membrane that permits exchange of $CO_{2}$ when the plate is in the $CO_{2}$ incubator.

\subsection{MEA recordings}\label{recordings}
Extracellular voltage recordings of neuronal cultures were performed using an MEA 2100-System (Multichannel Systems, Reutlingen, Germany). Each MEA contained 120 electrodes with a 100 µm inter-electrode distance. All data were acquired at a 20 kHz sampling rate. All recordings were performed in culture media. The head stage temperature was set to 30°C with an external temperature controller, and the MEAs were equilibrated for 5 min on the head stage before data acquisition or after any pharmacological or temperature manipulation. Recording duration was 3 minutes. Only cultures at 14 days in vitro (DIV) or older were used for pharmacological experiments. 

\subsection{Data Processing}
Raw data was converted to HDF5 file format and processed offline. Spike detection was done with Matlab tools Waveclus~\cite{chaure2018novel}. Note that we did not apply spike sorting, since it may introduce considerable noise due to unsupervised clustering methods when trying to obtain the neuron (or unit) information~\cite{chen2013overview}, and there are many different spike sorting algorithms~\cite{chaure2018novel,chung2017fully}, which give different outputs. Extracellular voltage recordings were bandpass filtered using cutoff frequencies of 200Hz and 4000Hz. Only negative deflections in the voltage records were labelled as spikes when the amplitude exceeded 6 times the standard deviation of the median noise level. Spike times and amplitudes were recorded and used for downstream analysis.

\section{Probabilistic Model}\label{model}
\begin{figure}[ht]
  \centering
  \includegraphics[width=\linewidth]{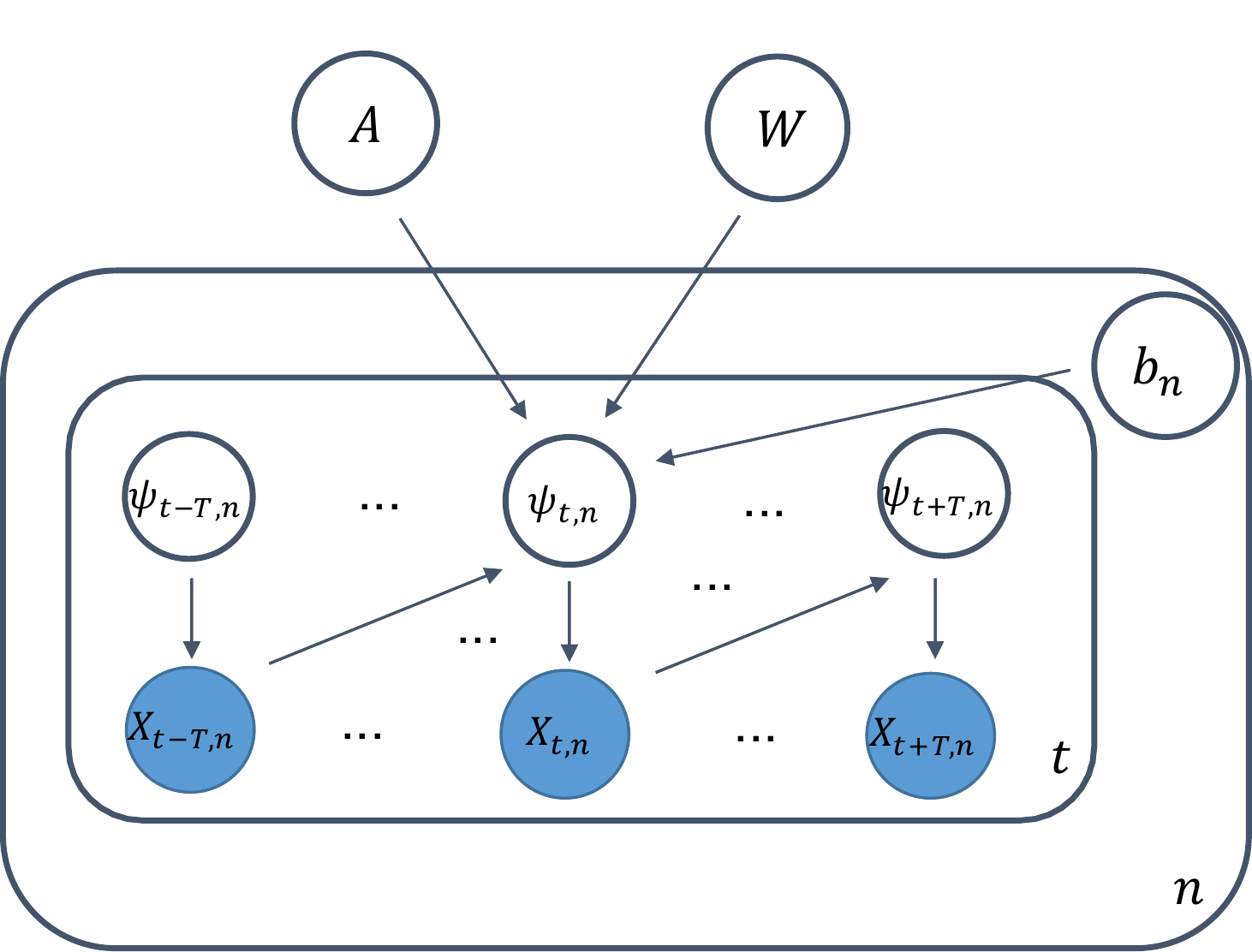}
  \caption{Probabilistic graphical model for MEA. The model describes how the underlying connectivity of a network of electrodes ($A$ and $W$) can lead to the observed spike trains $X_{t,n}$. }
  \Description{model}
  \label{fig:model}
\end{figure}

In this section, we briefly review the probabilistic model of neuronal spike trains introduced in \cite{Linderman} along with our choice of parameterization. Table~\ref{tab:notation} summarizes some common notations that we will use in this paper.  At a high level, the model describes how the underlying connectivity network affects the activation propensity of each electrode over time, producing the observed spike firing pattern measured over the entire MEA. Specifically, the model is composed of three parts: a network model specifying the underlying connectivity of the electrodes, an activation propensity model detailing how a network along with past spike history affects the probability of a spike at a time bin, and a spiking observation model mapping the activation propensity to the observed binary spike trains. Note that an electrode can fire no more than once in one time bin because of the refractory period in neurons. A probabilistic graphical model of this is shown in Fig~\ref{fig:model}.

\begin{table}
  \caption{Notations}
  \label{tab:Notations}
  \begin{tabular}{ccl}
    \toprule
    Notations&Description\\
    \midrule
    $X_{t,n}$ & The observed spike at time bin $t$ for electrode $n$\\ 
    $A$ & Adjacency matrix\\
    $W$ & Weight matrix\\
    $b_n$ &  The baseline activation of electrode $n$\\
    $N$ & Number of electrodes\\
    $T$ & Autoregressive window of influence\\
    $\psi_{t,n}$ & The activation of electrode $n$ at time bin $t$\\
    $N_o$ & Number of overlapped electrodes when split\\
    $W_o$ & Weight matrix of the overlapped region\\
    $N_s$ & Sample number of Bayesian inference\\
    $\rho$ & The prior for connection probability\\
    $\mu_{w_n}$ & Mean for the $n$th row of $W$\\
    $\mu_{b}$ & Mean of the bias vector\\
    $S_{w_n}$ & Covariance for the $n$th row of $W$\\
    $S_{b}$ & Covariance of the bias vector \\

  \bottomrule
\end{tabular}
\label{tab:notation}
\end{table}

\subsection{Network Model}
The network model aims to capture the key properties of the underlying functional network of the electrode population. Specifically, it seeks to represent that those connections are potentially directional and different in strength.
To accommodate this, a weighted directed graph is used where the edge weights represent the strength of the connection between two electrodes.  This is incorporated as two matrix-valued latent variables $A \in \{0,1\}^{N \times N}$ and $W \in \mathbb{R}^{N \times N}$ corresponding to a binary adjacency matrix and a real-valued weight matrix respectively. 

\subsection{Activation Propensity Model}
A neuron can either fire spontaneously or as a response to communications (spikes) it receives from incoming, connected neurons.  Given a particular realization of the electrode network, $A$ and $W$, the instantaneous activation of electrode $n$ at time bin $t$, $\psi_{t,n}$ is modeled as a linear, autoregressive function of the lagged spikes from neighboring electrodes:
\begin{equation}
    \psi_{t,n} = b_n + \sum_{m = 1}^N \sum_{\Delta t = 1}^{T} A_{m\rightarrow n} W_{m\rightarrow n} e^{-\Delta t/\tau} X_{t - \Delta t,m},
    \label{equ_3.2}
\end{equation}
Here, $b_n$ represents the baseline activation rate for electrode $n$ in the absence of influence from any other electrode. $A_{m\rightarrow n} \in \{0, 1\}$ is a binary variable indicating whether or not there exist directed connections from electrode $m$ to electrode $n$. The weight $W_{m\rightarrow n}$ is the connection strength from electrode $m$ to electrode $n$.  The activation rate $\psi$ is linearly adjusted by the lagged spikes from neighboring electrodes. The strength of the lagged spike is weighted by the strength of the connection to the neighbor and an exponentially decreasing function of time, inspired by the synapse connectivity measurement in ~\cite{bio-inspired}, with time constant of $\tau = 15ms$. This prioritizes recent spikes from strongly connected neighbors. We consider both positive and negative $W_{m\rightarrow n}$, which captures that neuronal connections may be excitatory or inhibitory in nature.   It is possible for a spike to decrease the propensity of firing when a weight is negative. 

\subsection{Bernoulli Observation Model}
Our spike train data consists of binary observations of whether electrode $n$ fired at time bin $t$, $X_{t,n}$.  This is modeled as a Bernoulli random variable with probability dependent on the activation propensity, $\sigma(\psi_{t,n}) = e^{\psi_{t,n}}(1+e^{\psi_{t,n}})^{-1}$, where $\sigma$ is the logistic function that maps the propensity to a probability.

\section{Bayesian Inference}\label{Bayesian}
Based on the previous section, the full model, including the priors, is as follows:  

\begin{equation}
\begin{aligned}
    A_{i,j} &\sim  \operatorname{Bernoulli}(\rho)\\
    \{\mu_{w_n}, \mu_{b}\}, {S_{w_n},S_{b}} &\sim \operatorname{Normal-Inverse-Wishart}(0, 1, \mathbf{I}, 3) \\
    w_n | \mu_{w_n}, S_{w_n} &\sim \operatorname{Normal}(\mu_{w_n}, S_{w_n}) \\
    b | \mu_{b}, S_{b} &\sim \operatorname{Normal}(\mu_{b}, S_{b}) \\
    \psi_{t,n} &= b_n + \sum_{m = 1}^N \sum_{\Delta t = 1}^{T} A_{m\rightarrow n} W_{m\rightarrow n} e^{-\Delta t/\tau} s_{t - \Delta t,m} \\
    X_{t,n} &\sim \operatorname{Bernoulli}(\sigma(\psi_{t,n})),
\end{aligned}
\end{equation}
where $w_n$ denotes the $n$th row of the matrix $W$. The hyper-parameter $\rho$ affects the prior over the connectivity matrix $A$.

We apply an efficient Gibbs sampler, which exhibits scalable parallelism, derived from~\cite{Linderman}. Sampling the posterior over the discrete adjacency matrix $A$ is the most challenging step.  Due to conjugacy, we can integrate over $W$ and sample $A$ from its collapsed conditional distribution. We update $A$ and $W$ by collapsing out $W$ to directly sample each of $A$’s elements. We iterate over each $A_{m\rightarrow n}$ and sample it from its conditional distribution. After that, we sample $W$ from its conditional Gaussian likelihood.

We employ a novel splitting strategy to make the inference method amenable to high density arrays.  We present and verify the methodology in the following sections.

\section{split}\label{split}

\begin{figure}[ht]
  \centering
  \includegraphics[width=\linewidth]{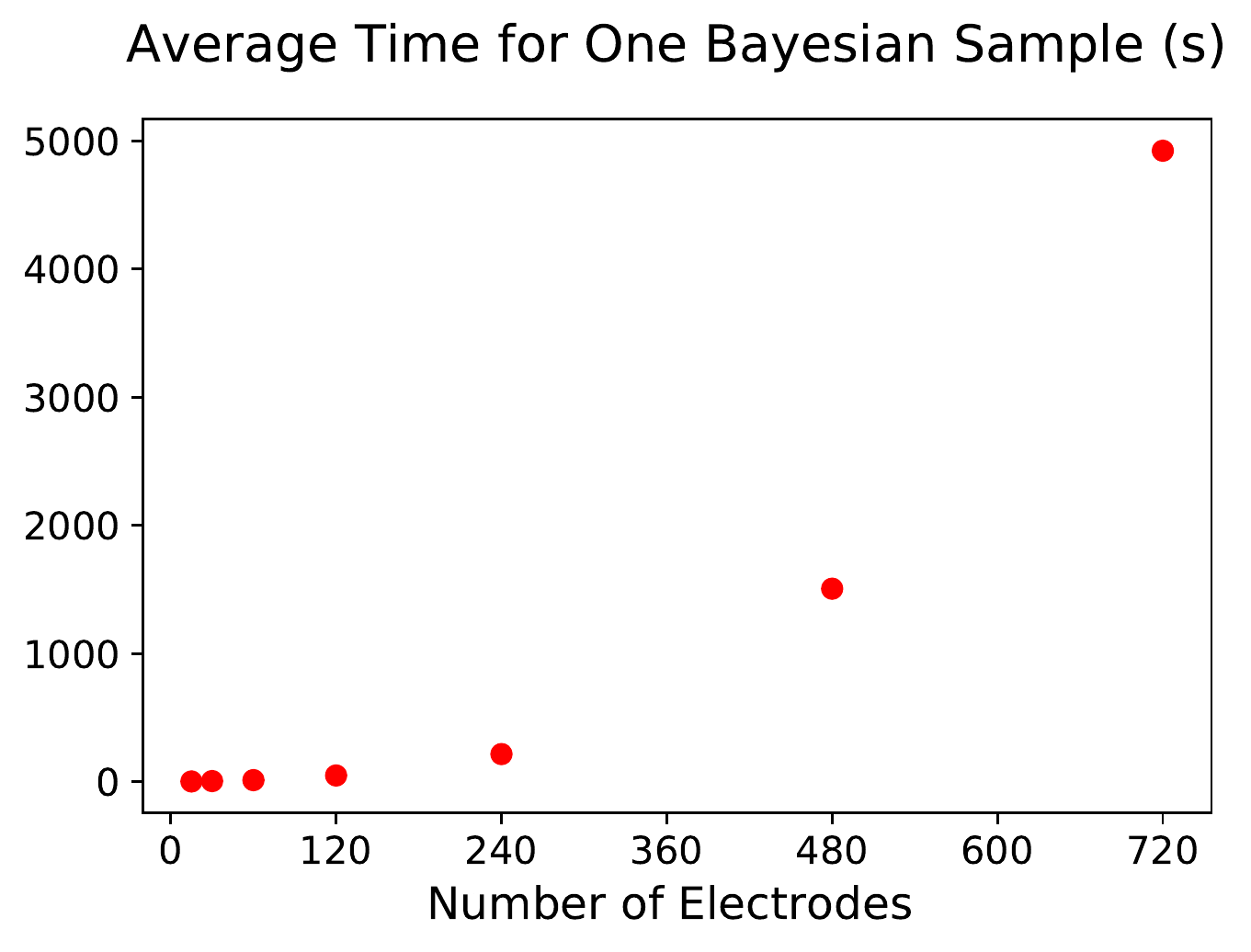}
  \caption{Average time for one Bayesian sampling exhibits quadratic growth according to the number of electrodes. The time reported is the sampling time computed in parallel with 24 central processing units (CPU) cores.}
  \Description{Time}
  \label{fig:Time}
\end{figure}
\begin{figure}[ht]
  \centering
  \includegraphics[width=\linewidth]{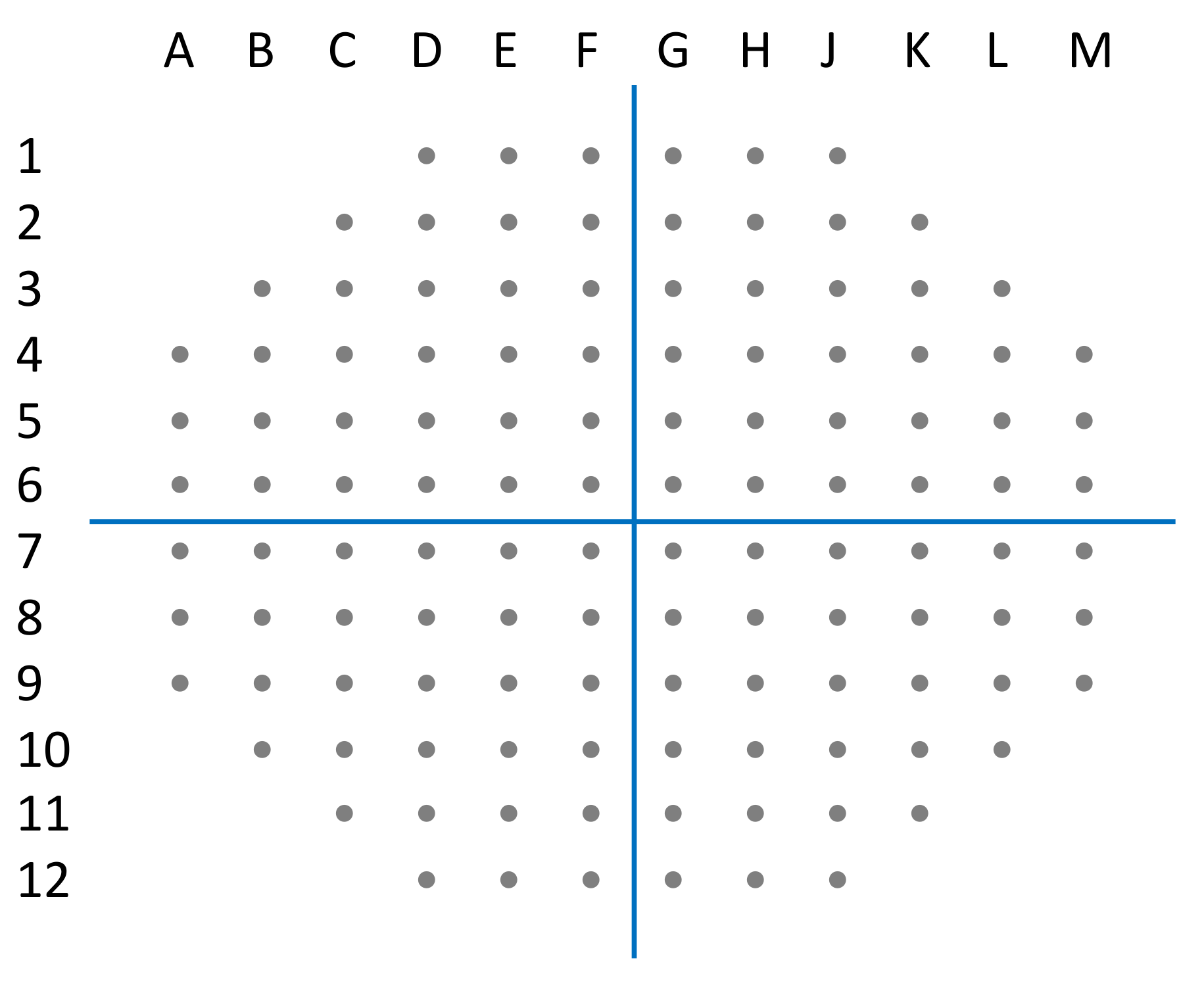}
  \caption{Non-overlapping split. We split the large array into four regions. For the first level, we infer individually on the smaller regions. For the second level, we treat each whole region as one hidden super node and infer the regional connection strength between each pair of sub-regions.}
  \Description{Non-Overlap Split}
  \label{fig:split}
\end{figure}
\begin{figure}[ht]
  \centering
  \includegraphics[width=\linewidth]{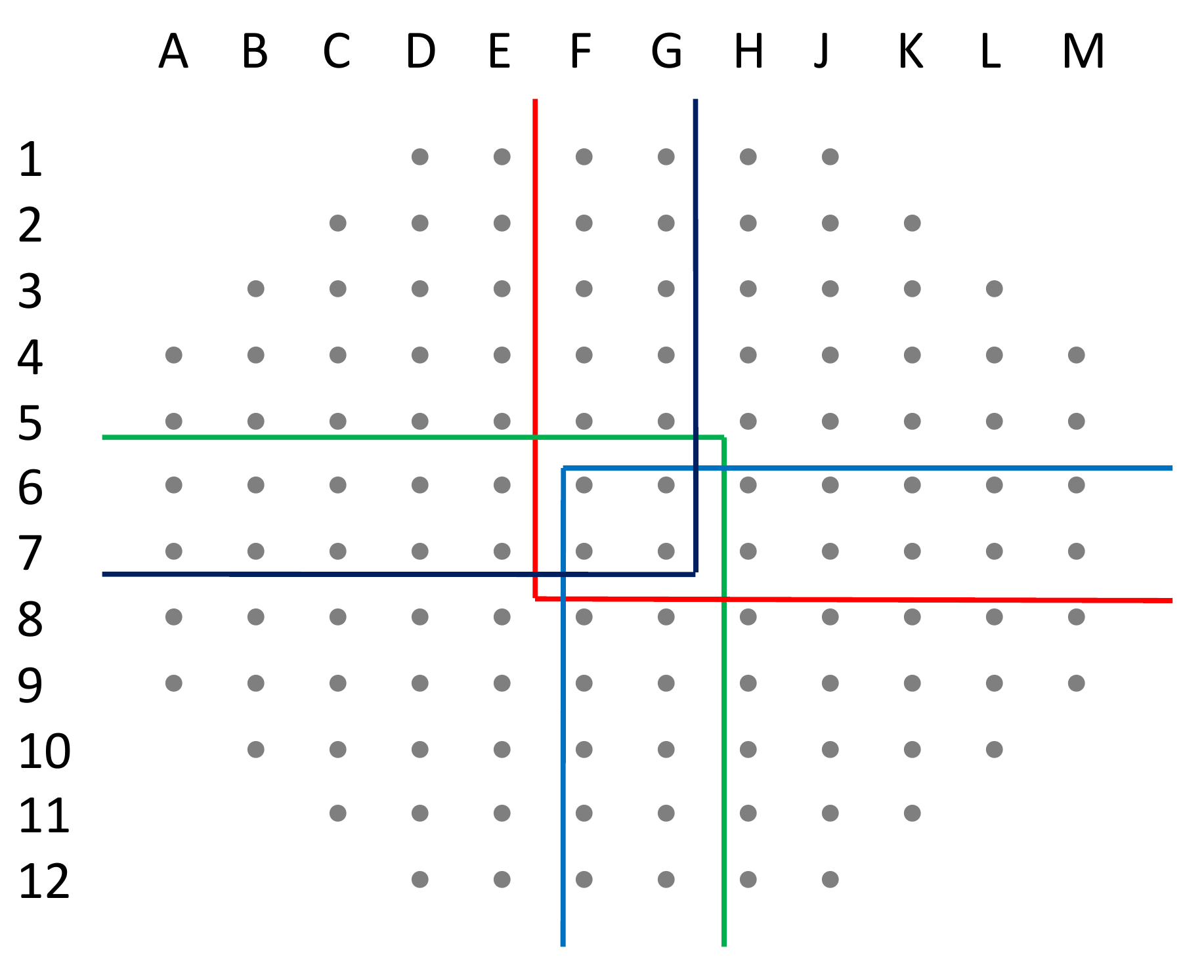}
  \caption{Overlapping split. To better capture the connections on the border of the regions, an overlapping split is used.}
  \Description{Overlap Split}
  \label{fig:split2}
\end{figure}

We have 120 electrodes in the MEA device in this paper. However, there are 26,400 electrodes in the MaxWell complementary metal oxide semiconductor (CMOS) MEA device.  As we can see in Fig.~\ref{fig:Time}, the average time for one sample increases drastically with the size of the array. That is because the dimensions of parameters ($A$ and $W$) to be estimated exhibit quadratic growth according to the number of electrodes. The time reported in Fig.~\ref{fig:Time} is the sampling time, computed in parallel with 24 CPU cores. To deal with the time complexity challenge, we propose a hierarchical inference procedure. As shown in Fig.~\ref{fig:split}, we split the whole large array into fixed number (for example, 4) of regions. In the first level of the algorithm, we perform Bayesian inference individually on each of the smaller regions.   In the second level, we treat each whole region as one group by taking the mean of all the spike trains in the region. We apply the same probabilistic model to infer the regional connectivity, which is the connection strength between each pair of regions. This heuristic splitting strategy is inspired by the biological phenomenon of regional connections in brains. The splitting strategy decreases the average sampling time quadratically in accordance with the number of sub-regions.  To get a better understanding of the connections on the border of any two regions, we further propose an overlapping split mechanism, as shown in Fig.~\ref{fig:split2}. 

\section{Results on synthetic data}\label{synthetic}

Due to the lack of ground truth functional connectivity in an MEA, we first use synthetic data to check the effectiveness of the probabilistic model and our splitting strategies. We use the ground truth model to generate synthetic data and compare the functional connectivity inferred from synthetic data with the ground truth. For all the experiments in this section and the next section, we use spike trains with shape of 180, 000 * 120. We apply parallel sampling with 24 CPUs. The Gibbs sampler was run for 1000 iterations. The first 500 samples were discarded to account for burn-in. We verify that the chain has reached a steady state by observing that the parameter traces and the log-likelihood have converged.

\begin{figure}[ht]
    \centering
    \subfloat{\includegraphics[width=0.23\textwidth]{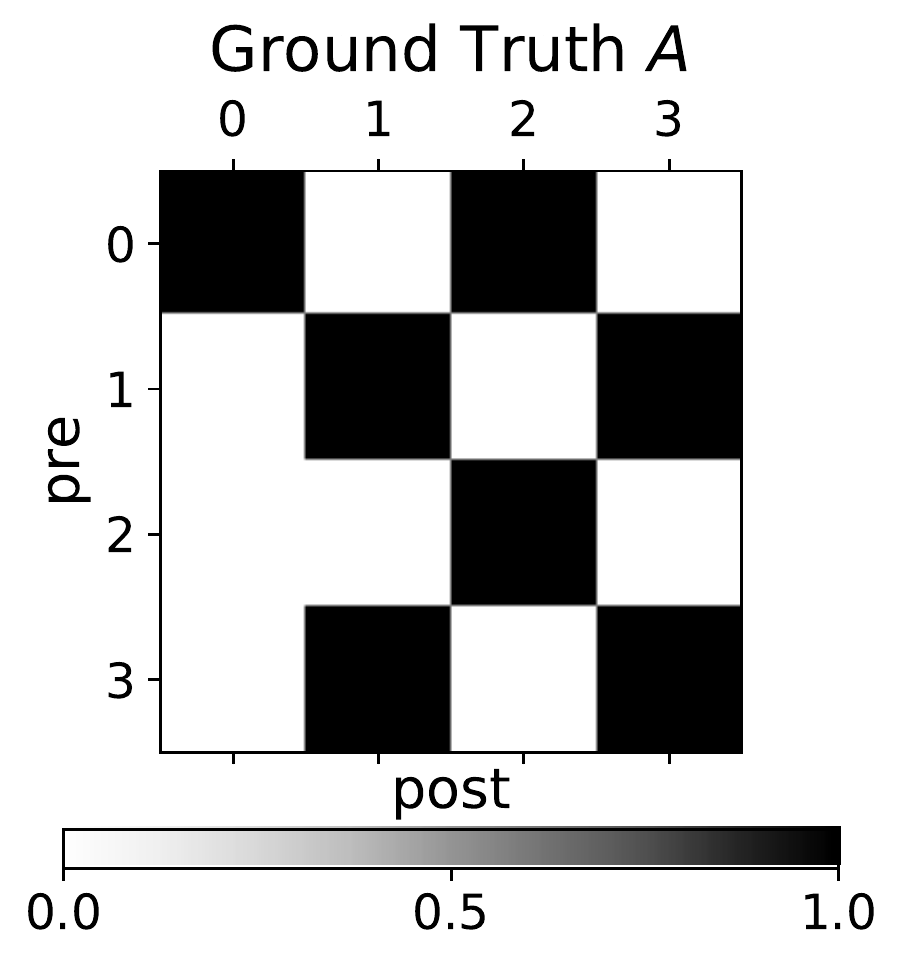}}
    \hfill
    \subfloat{\includegraphics[width=0.23\textwidth]{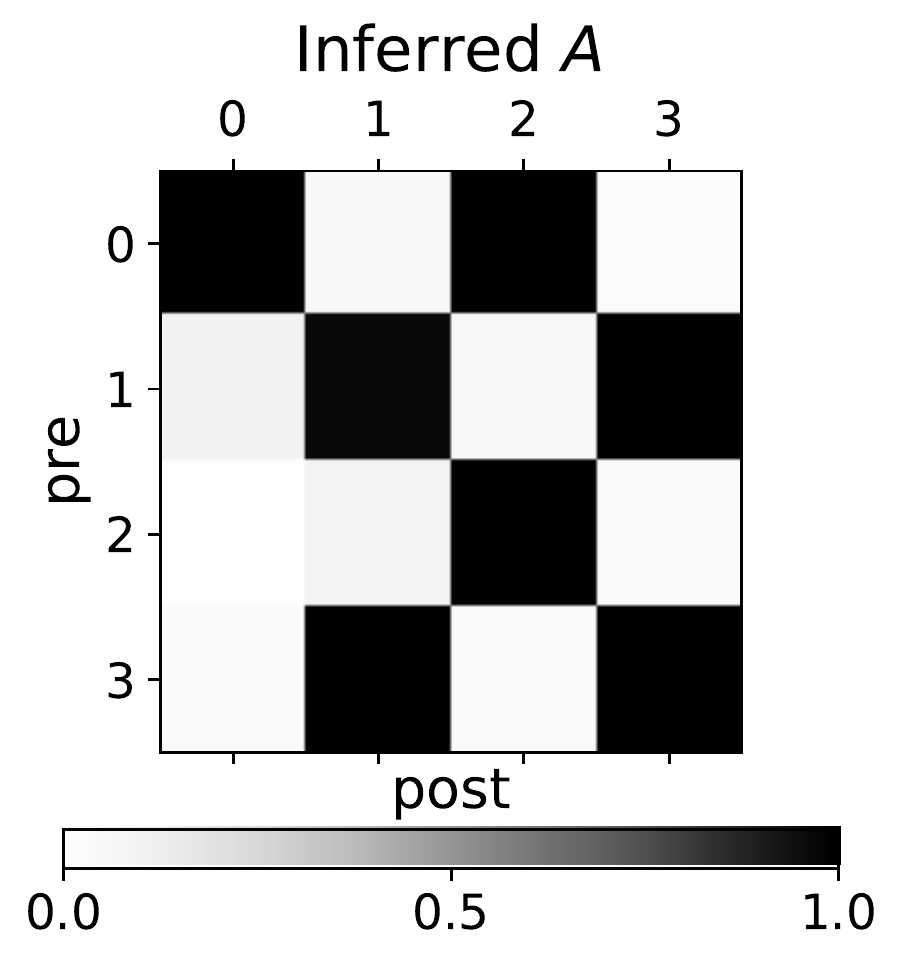}}
    \quad
    \subfloat{\includegraphics[width=0.23\textwidth]{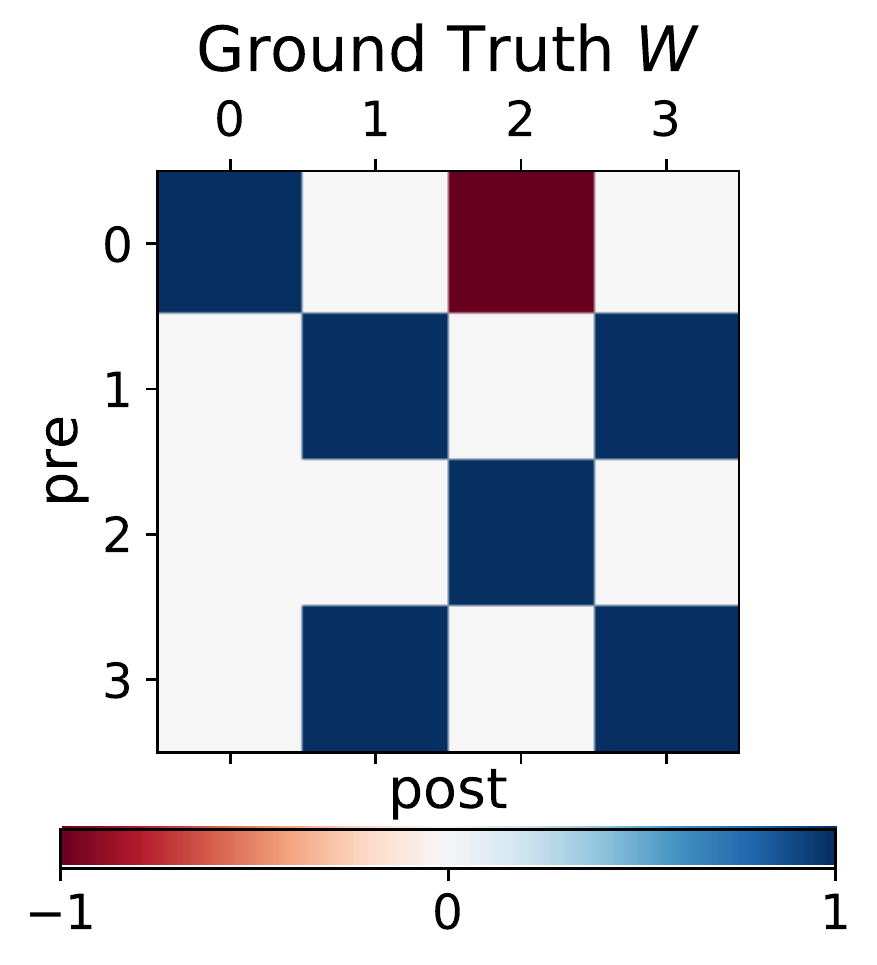}}
    \hfill
    \subfloat{\includegraphics[width=0.23\textwidth]{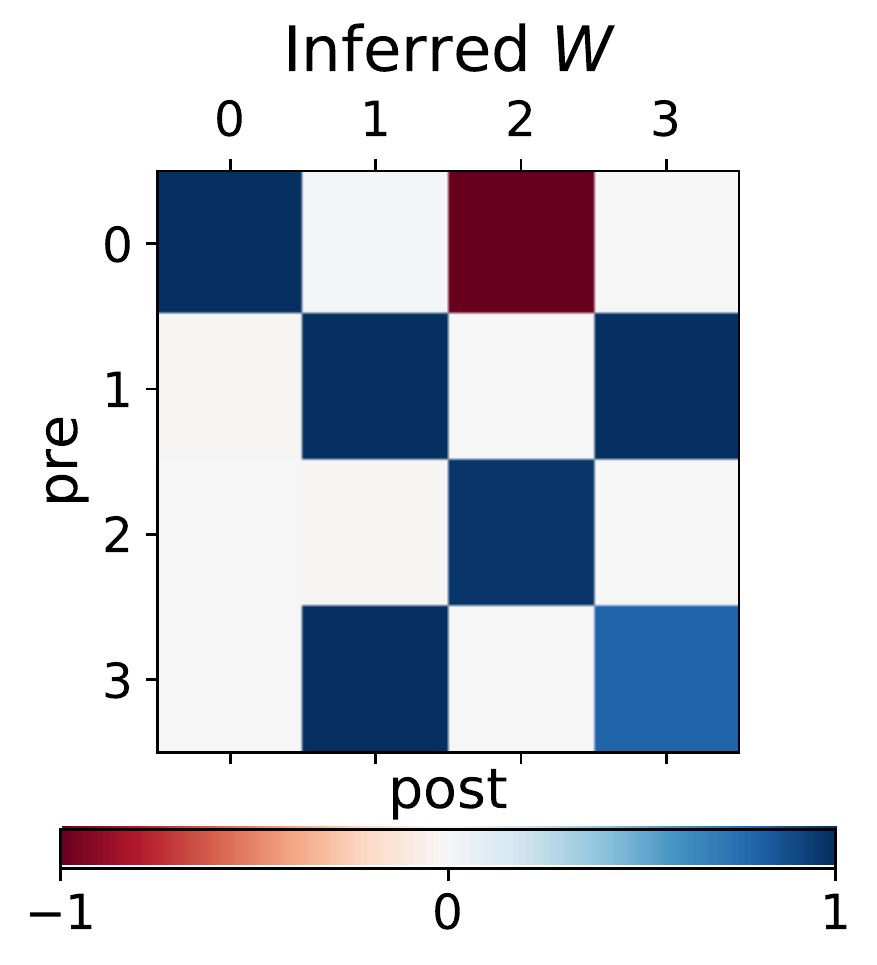}}
  \caption{Comparison of network inferred from synthetic data with 4 electrodes and ground truth. Zero indicates no connection. Minus one indicates an inhibitory connection, and one indicates an excitatory connection.
  Cosine similarity between the two $A$s is 0.99. Cosine similarity between the two $W$s is 0.99. }
  \Description{Comparison of synthetic data with 4 electrodes inferred network and ground truth. Cosine similarity between the two is 0.9658.}
  \label{fig:Synthetic4}
\end{figure}

\begin{figure}[ht]
    \centering
    \subfloat{\includegraphics[width=0.23\textwidth]{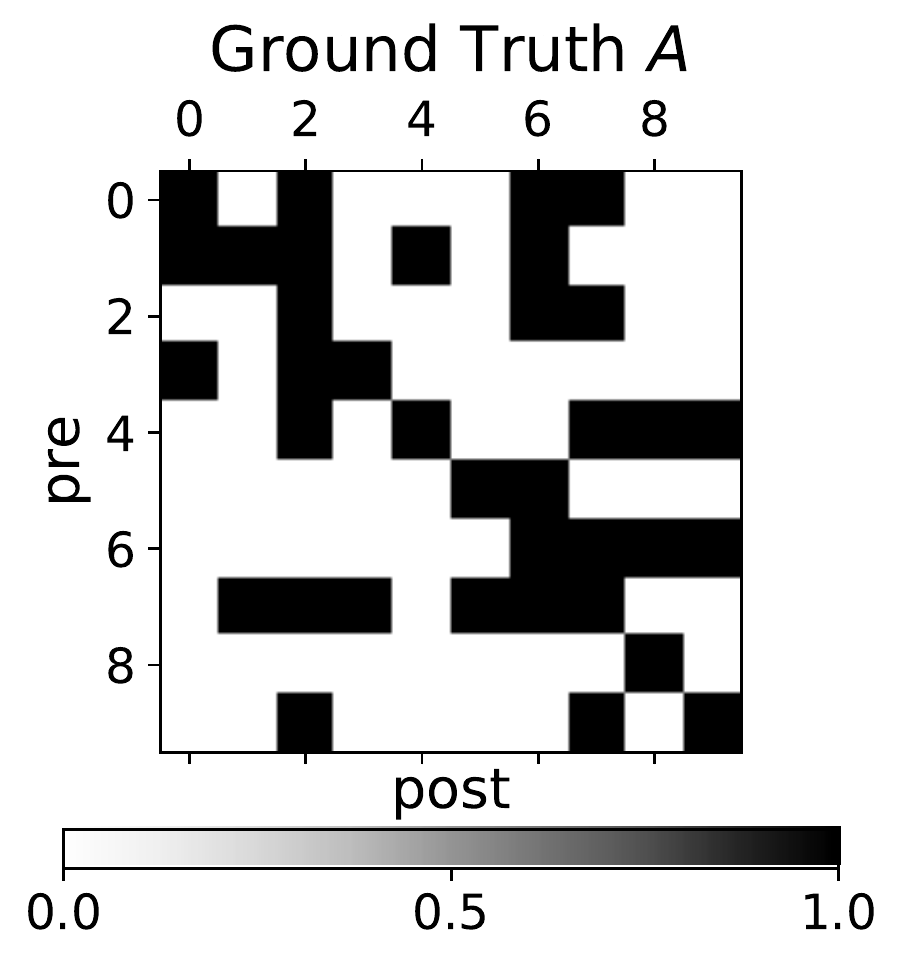}}
    \hfill
    \subfloat{\includegraphics[width=0.23\textwidth]{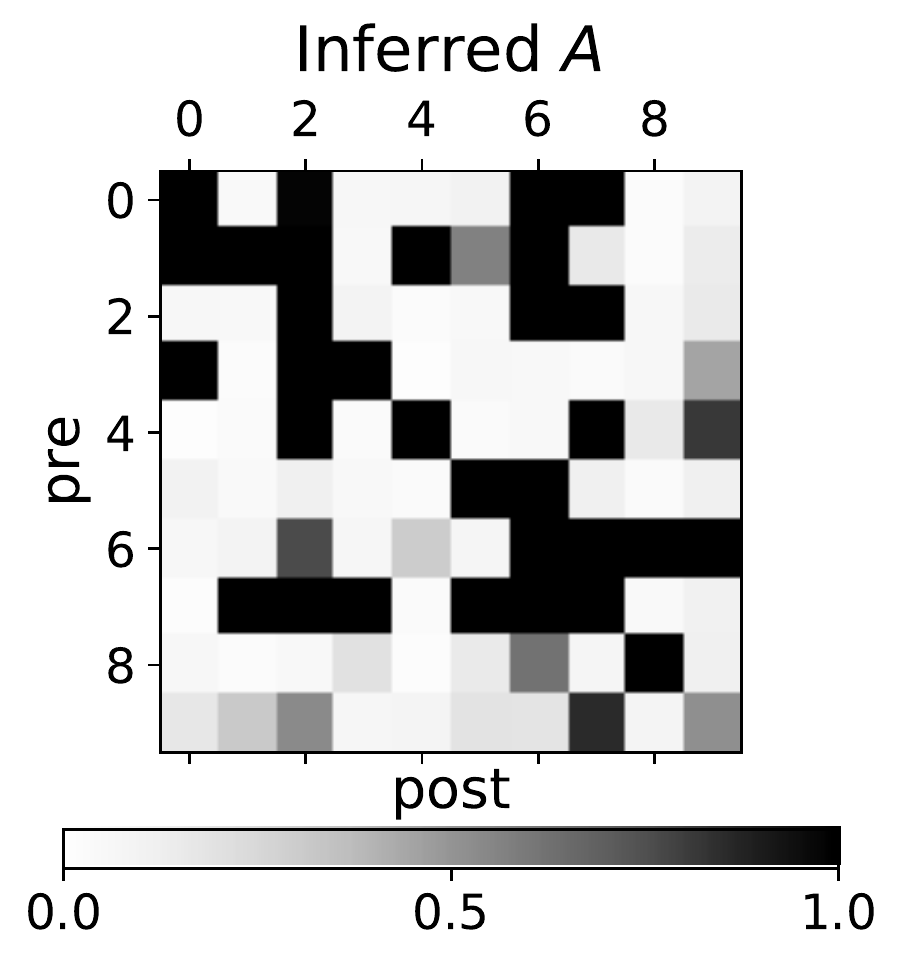}}
    \quad
    \subfloat{\includegraphics[width=0.23\textwidth]{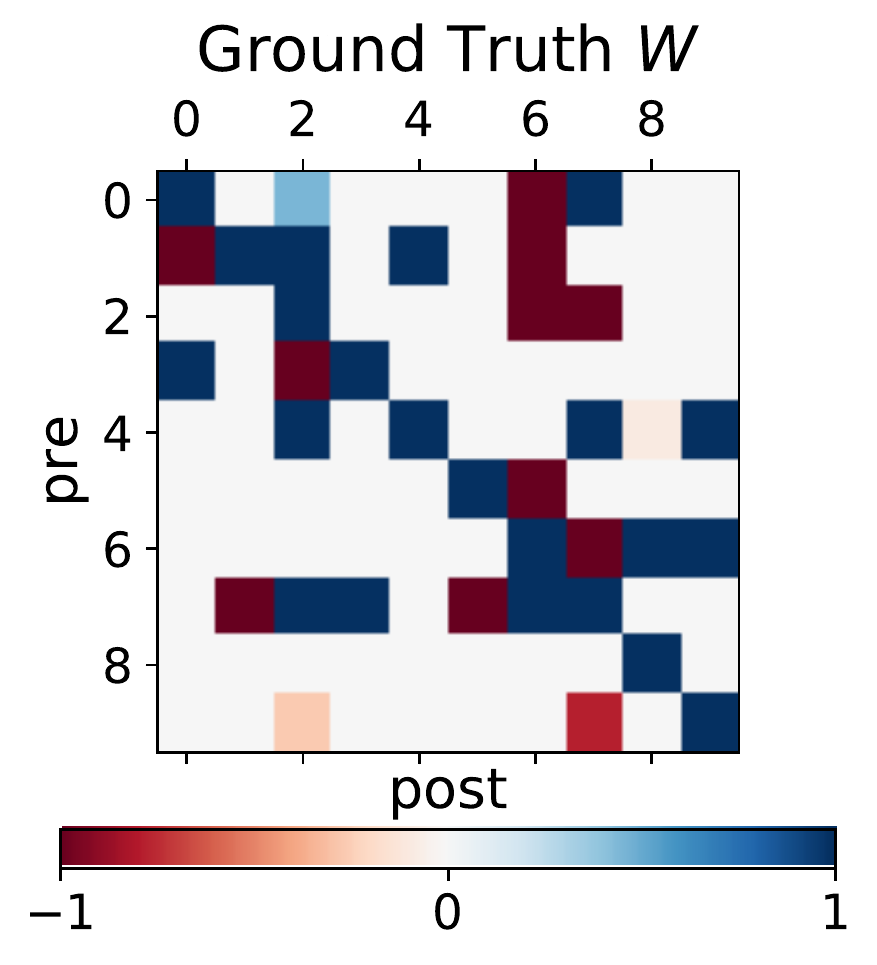}}
    \hfill
    \subfloat{\includegraphics[width=0.23\textwidth]{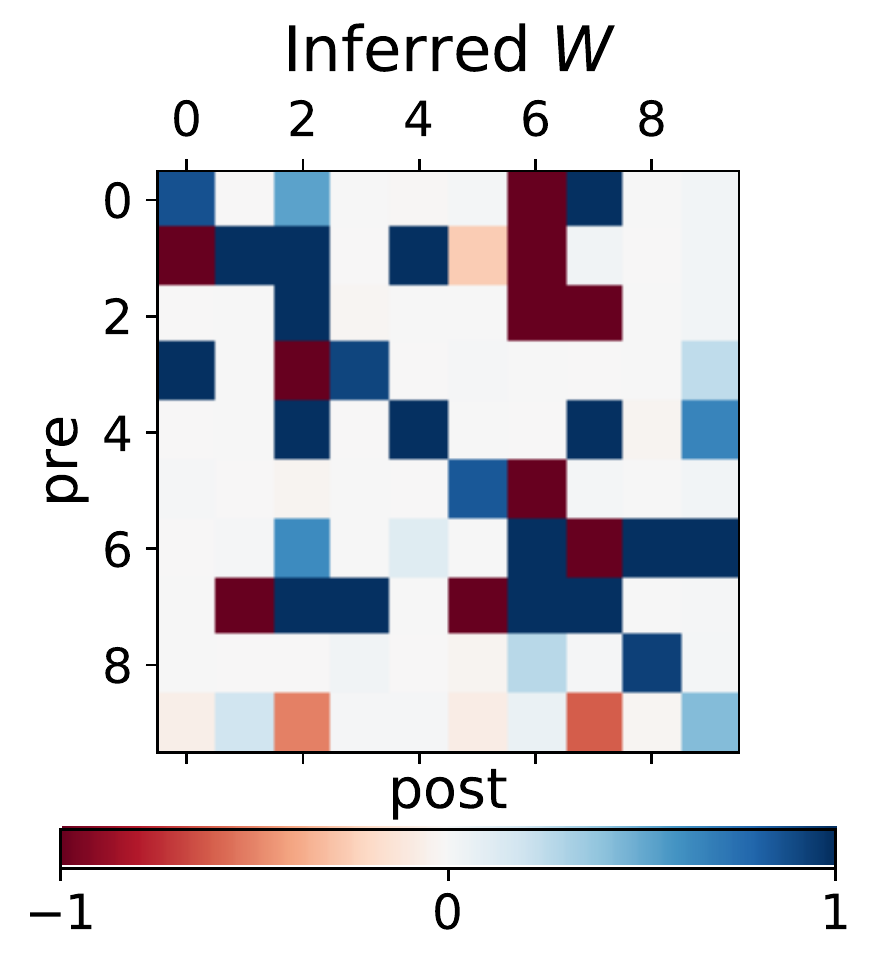}}
  \caption{Comparison of network inferred from synthetic data with 10 electrodes and ground truth. Cosine similarity between the two $A$s is 0.95. Cosine similarity between the two $W$s is 0.99.}
  \Description{Comparison of synthetic data with 10 electrodes inferred network and ground truth. Cosine similarity between the two $A$s is 0.9792. Cosine similarity between the two $W$s is 0.9975.}
  \label{fig:Synthetic10}
\end{figure}

In Fig.~\ref{fig:Synthetic4}, the inferred posterior mean is almost the same as the ground truth connectivity matrices. Cosine similarity measures the similarity between two matrices of an inner product space, which is widely used in high dimensional spaces~\cite{luo2018cosine}. The inferred functional network $A$ and $W$ both achieved a high cosine similarity of 0.99 compared with ground truth in this case. Similarly in Fig.~\ref{fig:Synthetic10}, we obtain cosine similarity of 0.95 and 0.99 for $A$ and $W$ when the number of electrodes increases to 10.

We also use synthetic datasets to verify the effectiveness of our splitting strategies. Here, we assume all the electrodes lie in a line. For the non-overlapping split, we use $N$ electrodes and split them into two sub-regions, "front" and "back", with equal size and apply the Bayesian inference on each region. We compared the separate split results with the Bayesian inference results without split and reported the cosine similarity in Tab.~\ref{tab:Non_overlap}. It is shown that the cosine similarities are consistently high with all different parameter settings, which validates the effectiveness of our splitting strategy. Similarly, for overlapping split, we test the results from split with $N_o$ overlapped electrodes. In Tab.~\ref{tab:overlap}, $N_{o}$ indicates the number of overlapped electrodes. $W_o$ is the weight matrix for the overlapped region. High cosine similarities in Tab.~\ref{tab:overlap} confirm the effectiveness of overlapping split strategy. From both Tab.~\ref{tab:Non_overlap} and Tab.~\ref{tab:overlap}, the network prior does affect the inference of $A$ and $W$ but in an indirect way, which is small in degree compared to the effect of the data.

\begin{figure}[ht]
    \centering
    \subfloat{\includegraphics[width=0.23\textwidth]{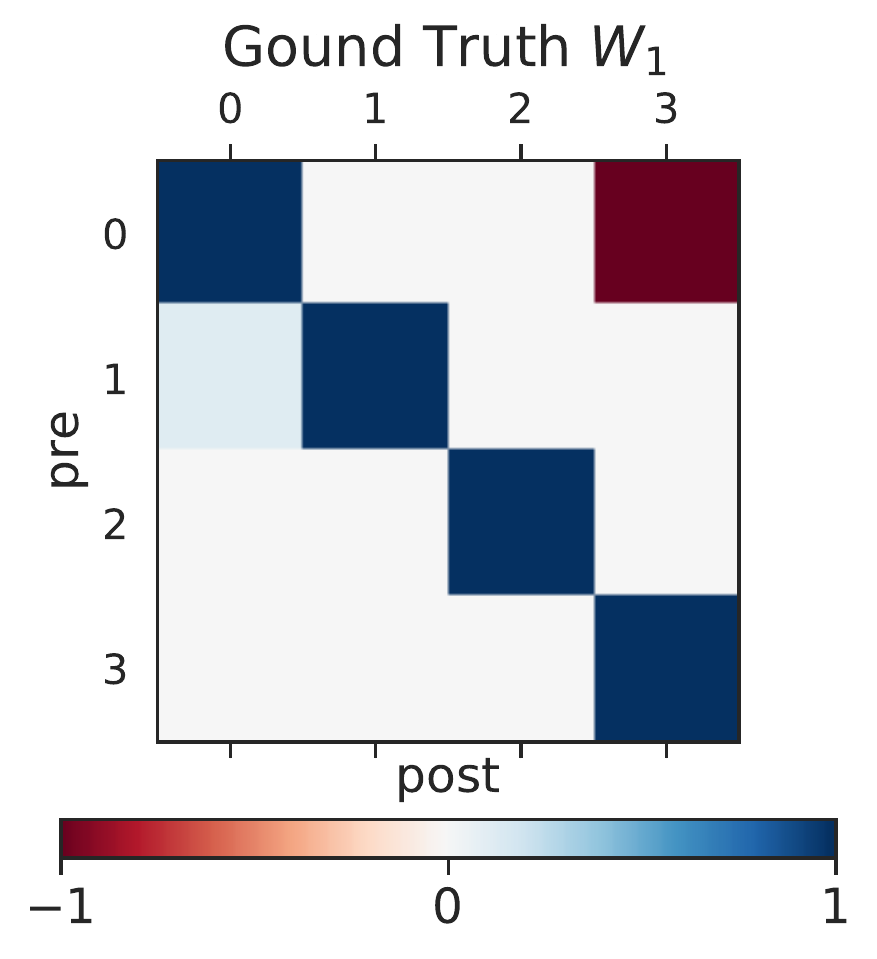}}
    \hfill
    \subfloat{\includegraphics[width=0.23\textwidth]{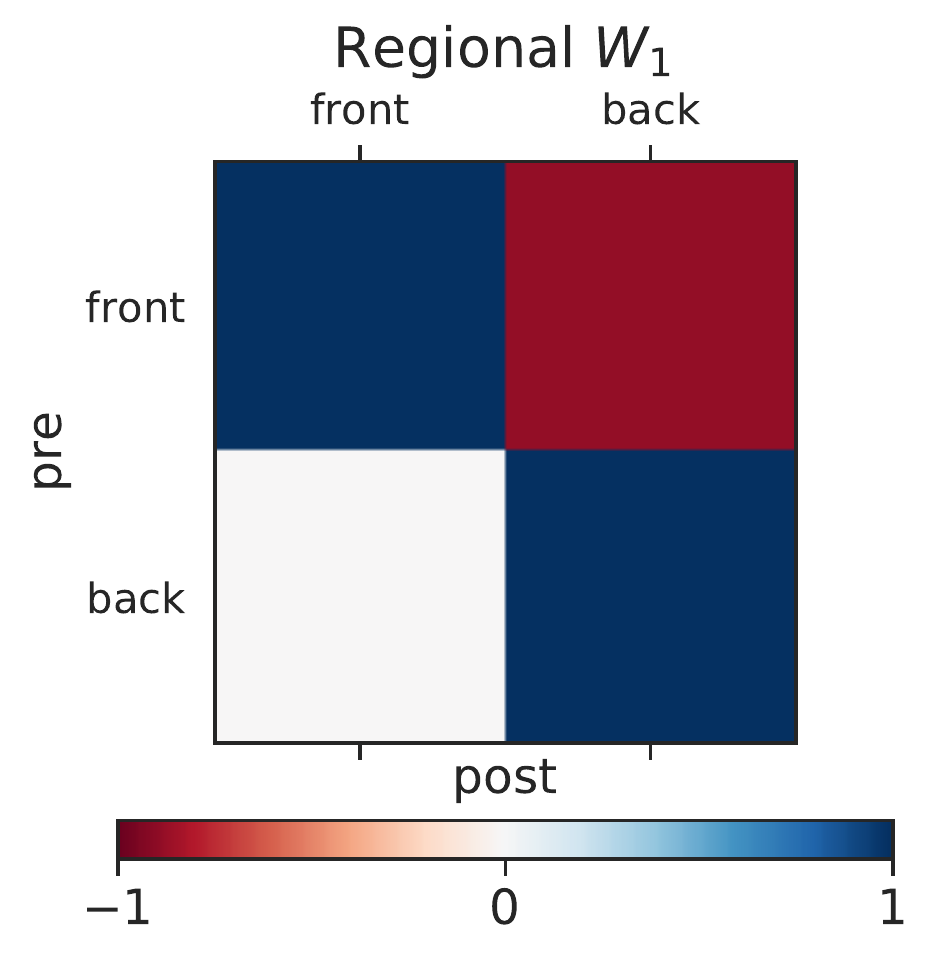}}
    \quad
    \subfloat{\includegraphics[width=0.23\textwidth]{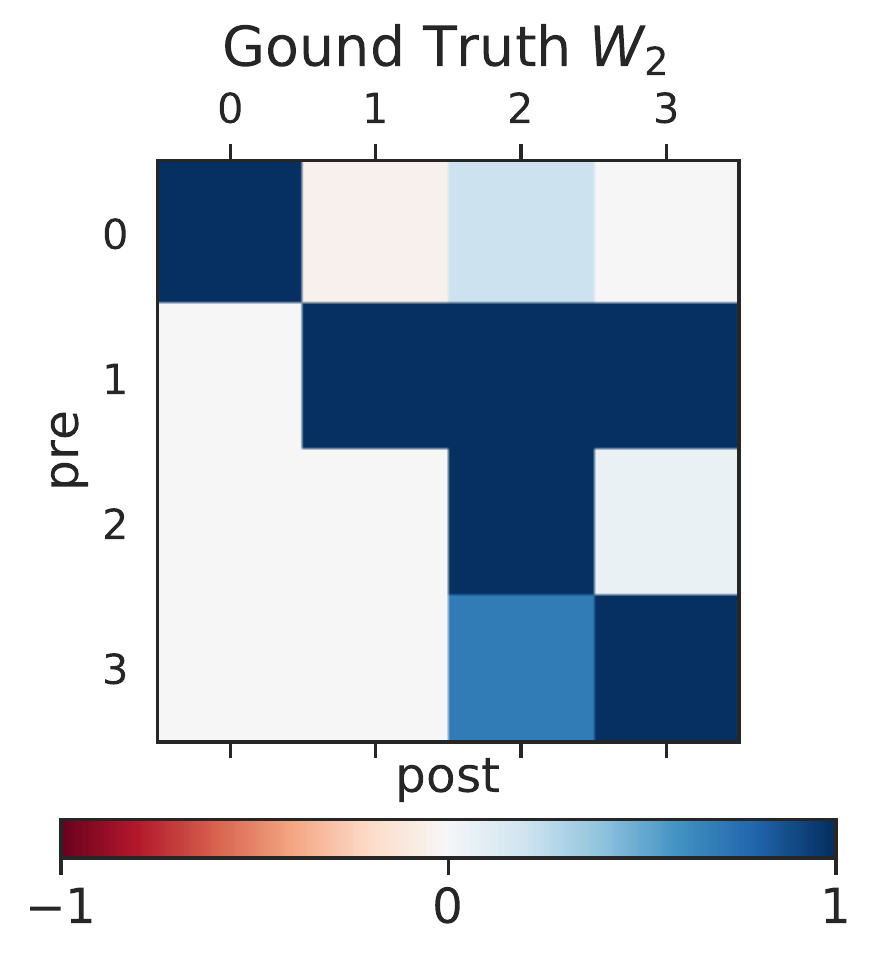}}
    \hfill
    \subfloat{\includegraphics[width=0.23\textwidth]{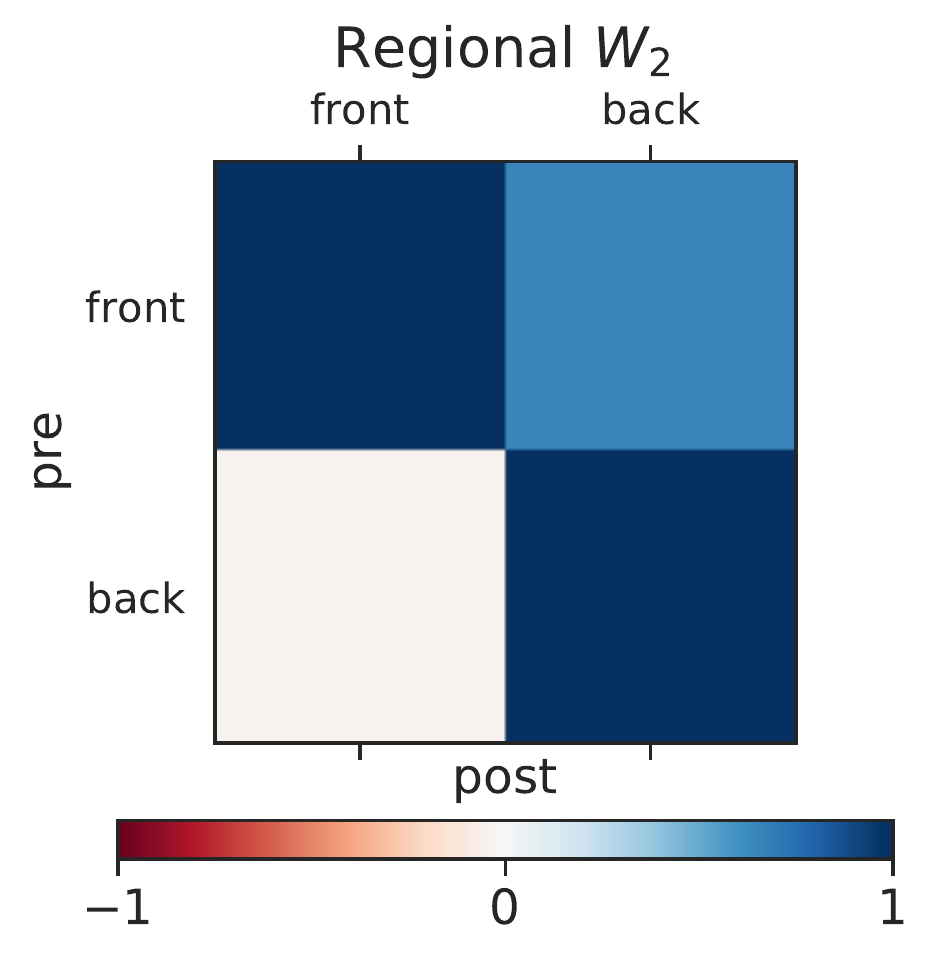}}
  \caption{Regional connections in non-overlapping split. "Front" indicates the region composed of electrodes 0 and 1 in the ground truth figure, while "back" indicates the region of electrode 2 and 3. 
  Each element in the inferred regional connectivity matrix $W$ reflects the elements of the corresponding regions in ground truth $W$ altogether. Regional inferences after split precisely reveal the overall connectivity between regions. }
  \Description{Regional connections in split}
  \label{fig:Synthetic2nd}
\end{figure}

Fig.~\ref{fig:Synthetic2nd} shows the effectiveness of regional connection inference after split. Each element in the inferred regional connectivity matrix $W$ summarizes the elements of the corresponding regions in ground truth $W$ altogether. Regional inferences after split precisely indicate the strength and the nature of the connections between regions.


\begin{table}
  \caption{Non-overlapping Results: Cosine Similarity}
  \label{tab:Non_overlap}
  \begin{tabular}{cccccccl}
    \toprule
    N&$S_w$&$\mu_b$&$\rho$&$W_{front}$&$A_{front}$&$W_{back}$&$A_{back}$\\
    \midrule
    10&1&0&0.5&0.93&0.99&0.94&0.98\\
    10&1&0&1&0.99&0.99&0.98&0.99\\
    10&1&5&0.5&0.99&0.99&0.98&0.99\\
    10&2&0&0.5&0.95&0.98&0.88&0.97\\
    20&1&0&0.5&0.91&0.99&0.95&0.99\\
    30&1&0&0.5&0.94&0.99&0.93&0.99\\
  \bottomrule
\end{tabular}
\end{table}

\begin{table}
  \caption{Overlapping Results: Cosine Similarity}
  \label{tab:overlap}
  \begin{tabular}{cccccccccl}
    \toprule
    N&$N_{o}$&$S_w$&$\mu_b$&$\rho$&$W_{front}$&$A_{front}$&$W_{back}$&$A_{back}$&$W_{o}$\\
    \midrule

    10&2&1&0&0.5&0.97&0.9&0.95&0.99&0.98\\
    10&2&1&0&1&0.99&0.99&0.99&0.99&0.99\\
    10&2&1&5&0.5&0.99&0.99&0.99&0.99&0.99\\
    10&4&2&0&0.5&0.92&0.98&0.95&0.99&0.93\\
    20&4&1&0&0.5&0.94&0.99&0.97&0.99&0.96\\
    30&4&1&0&0.5&0.90&0.99&0.93&0.99&0.97\\									
  \bottomrule
\end{tabular}
\end{table}

\section{results on real data}\label{real}
In this section, we apply our framework onto two sets of real MEA recordings. Throughout this section, we apply non-overlapping split with 4 regions of equal size as in Fig.~\ref{fig:split}. As we mentioned in Section~\ref{recordings}, we use 3-minute recordings, which present as spike trains with shape of 180, 000 * 120. For the Bayesian inference, we adjusted the prior hyper-parameters and verified that the results stayed consistent.

\subsection{Comparison with neuron experts' labeled ground truth}
\begin{figure}[ht]
  \centering
  \includegraphics[width=\linewidth]{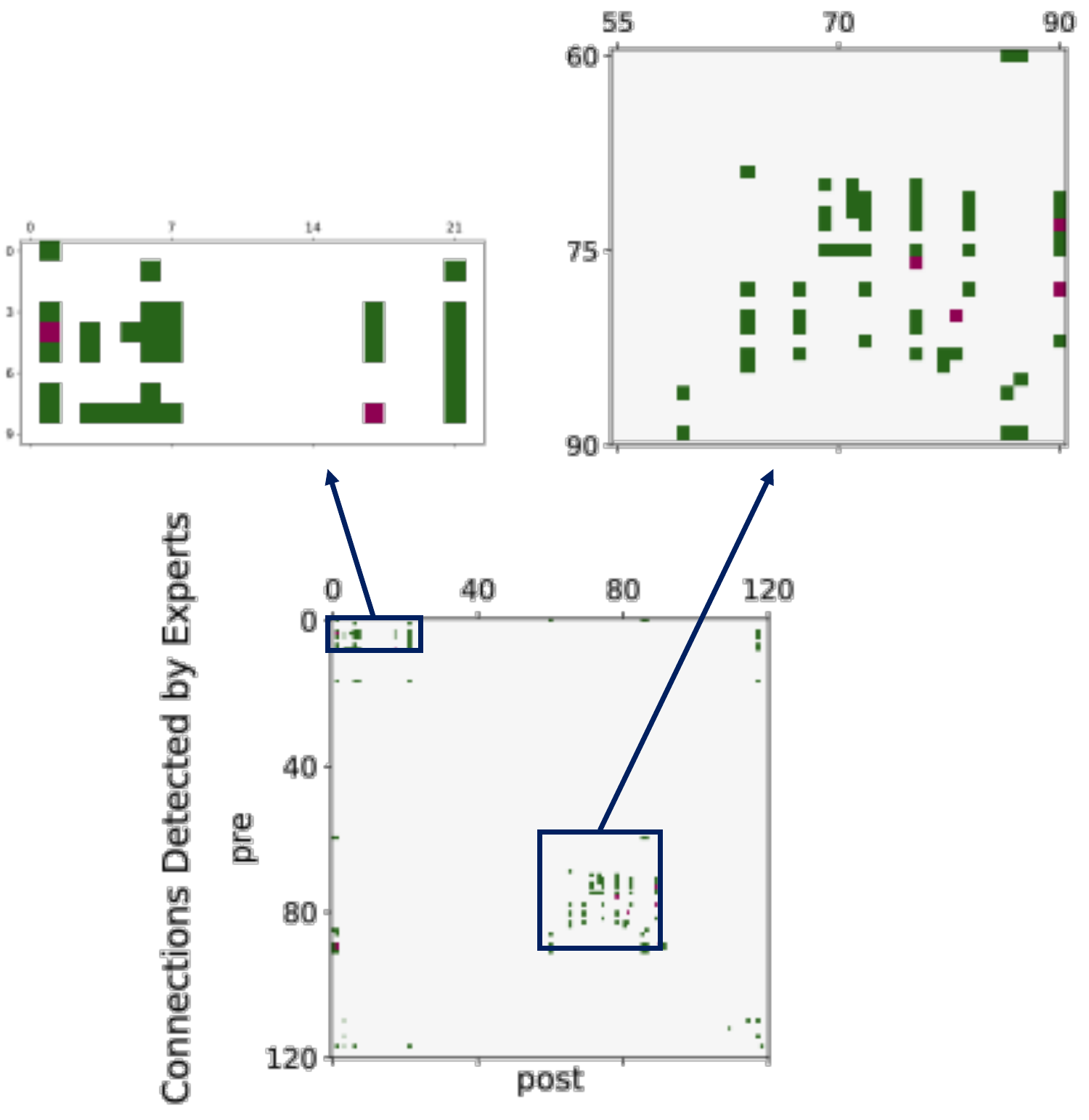}
  \caption{Comparison between our Bayesian inference and ground truth detected by neuroscience experts. Green represents the connections found by Bayesian inference, while red denotes the connections not found. 93.33\% of the connections detected by the neuroscience experts are detected by Bayesian inference.}
  \Description{Comparison between our Bayesian inference and Ground Truth detected by neuroscience expert. 93.33\% of the connections detected by biology expert are detected by Bayesian inference.}
  \label{fig:Compare}
\end{figure}

For the real dataset, since we don't have the ground truth, we adopted neuroscience experts' labelling as our reference. The neuroscience experts labelled those connections by watching MEA viewer~\cite{bridges2018mea}. We compare our Bayesian analysis with the expert labeled result. As is shown in Fig.~\ref{fig:Compare}, 93.33\% of all the connections detected by neuron experts are detected by our inference. We can see that the real recording has some regional patterns, which satisfies the intuition of our splitting strategy. Because of the difficulty in detecting those connections in neuroscience experiments, our neuroscience experts can find some connections that they are confident in but cannot guarantee to find all the connections. However, in our Bayesian inference, we gave the overall possible connections based on our probabilistic model.
Thus we found more connections using Bayesian inference compared with the neuroscience experts' result. We also calculated the latency between electrodes using CCG based on the neuroscience experts’ labeled ground truth connectivity. The latency shows a similar pattern as our inferred weights.

\subsection{Cadmium concentration vs. controlled experiment}

\begin{figure}[ht]
  \centering
  \includegraphics[width=\linewidth]{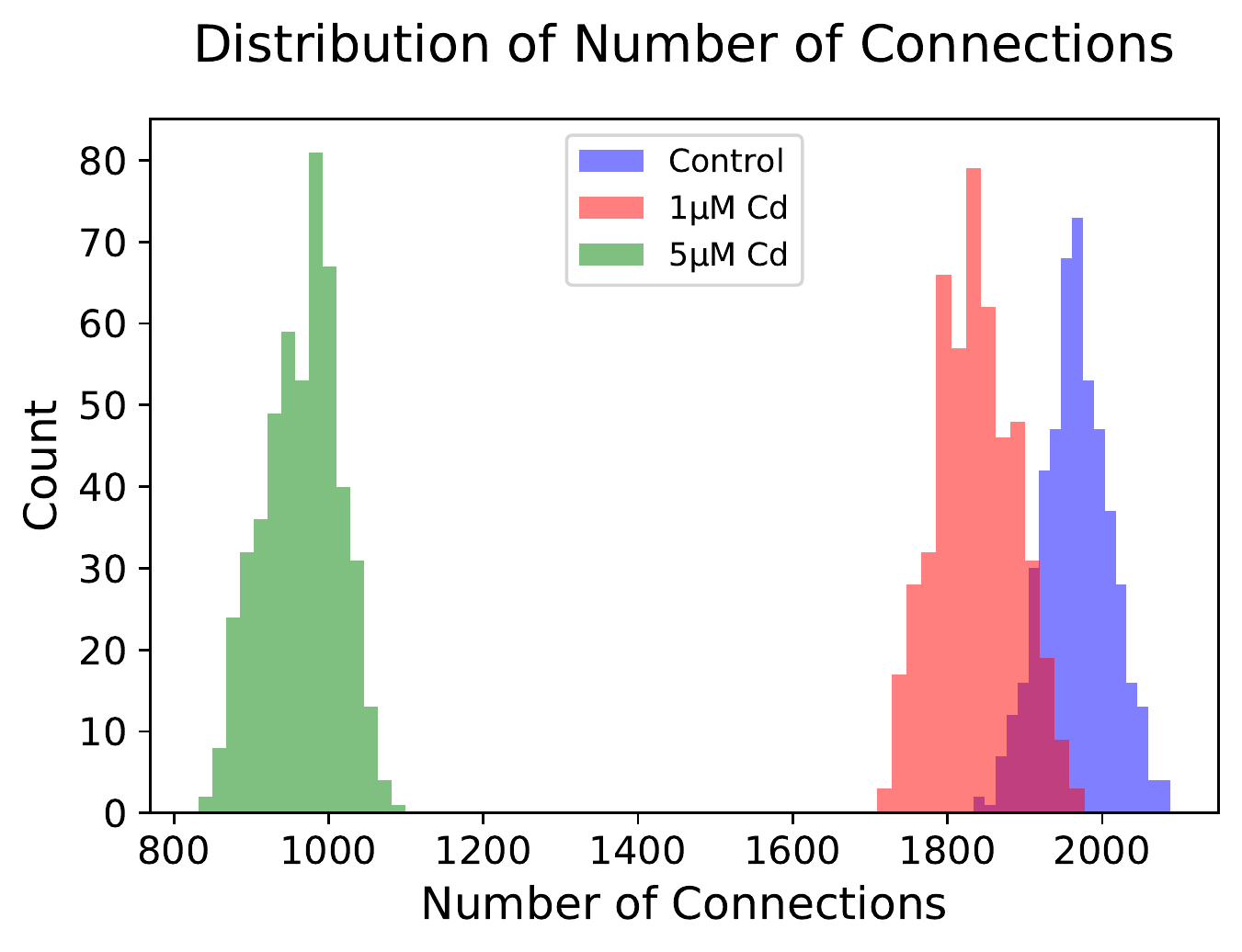}
  \caption{Posterior of number of connections with Cadmium concentration. Cd is short for Cadmium. The mean connection counts decrease with introduction of cadmium to the culture. Higher concentrations of cadmium lead to a further decrease in the number of connections in the functional networks.}
  \Description{Number of Connections Distribution with Cadmium Concentration}
  \label{fig:Distribution}
\end{figure}

\begin{figure}[ht]
  \centering
  \includegraphics[width=\linewidth]{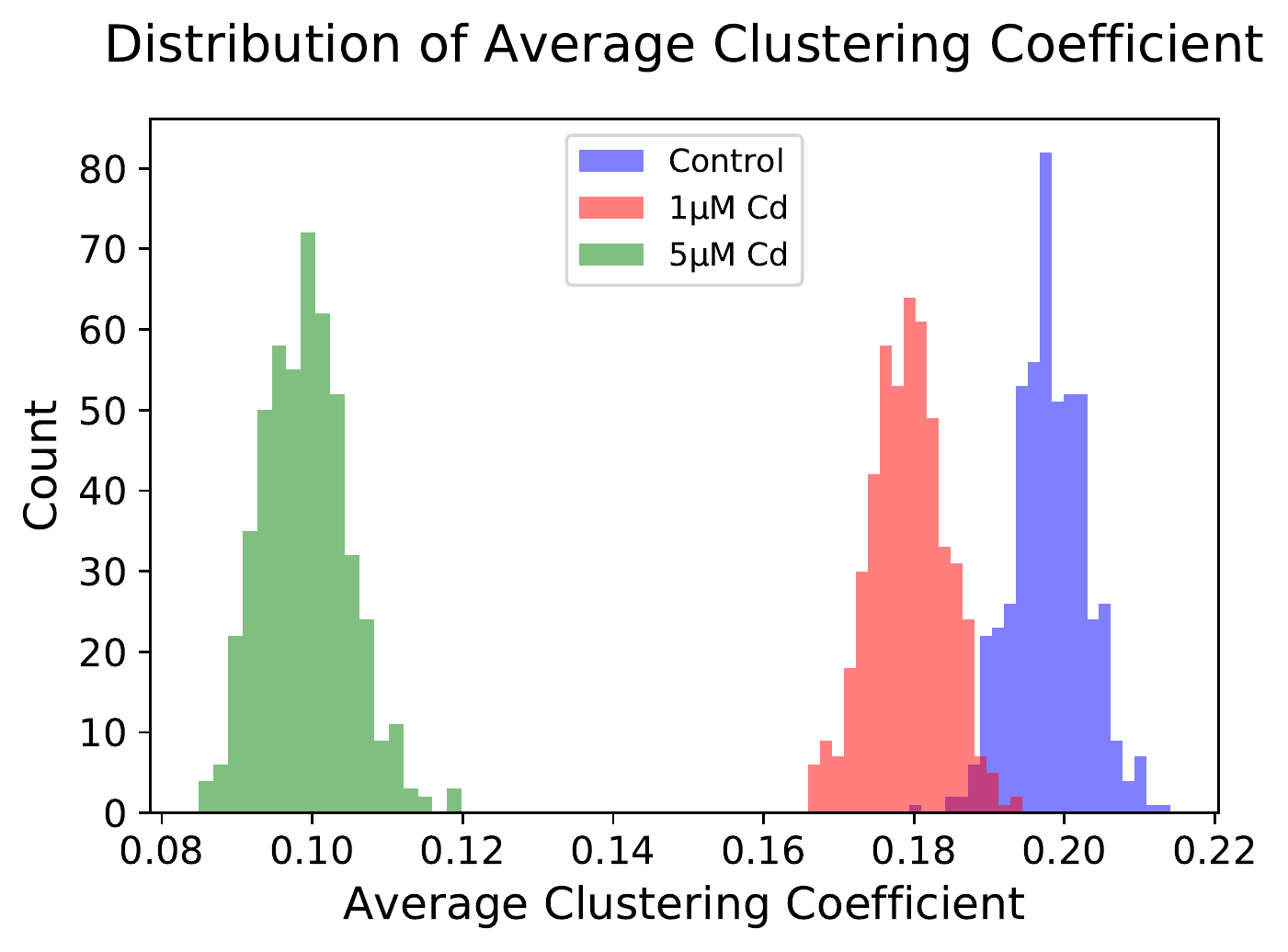}
  \caption{Posterior of average clustering coefficient with Cadmium concentration. Average clustering coefficient decreases with statistical significance with the introduction of Cadmium into the culture. }
  \Description{Number of Connections Distribution with Cadmium Concentration}
  \label{fig:Distribution_CC}
\end{figure}

\begin{figure}[ht]
  \centering
  \includegraphics[width=\linewidth]{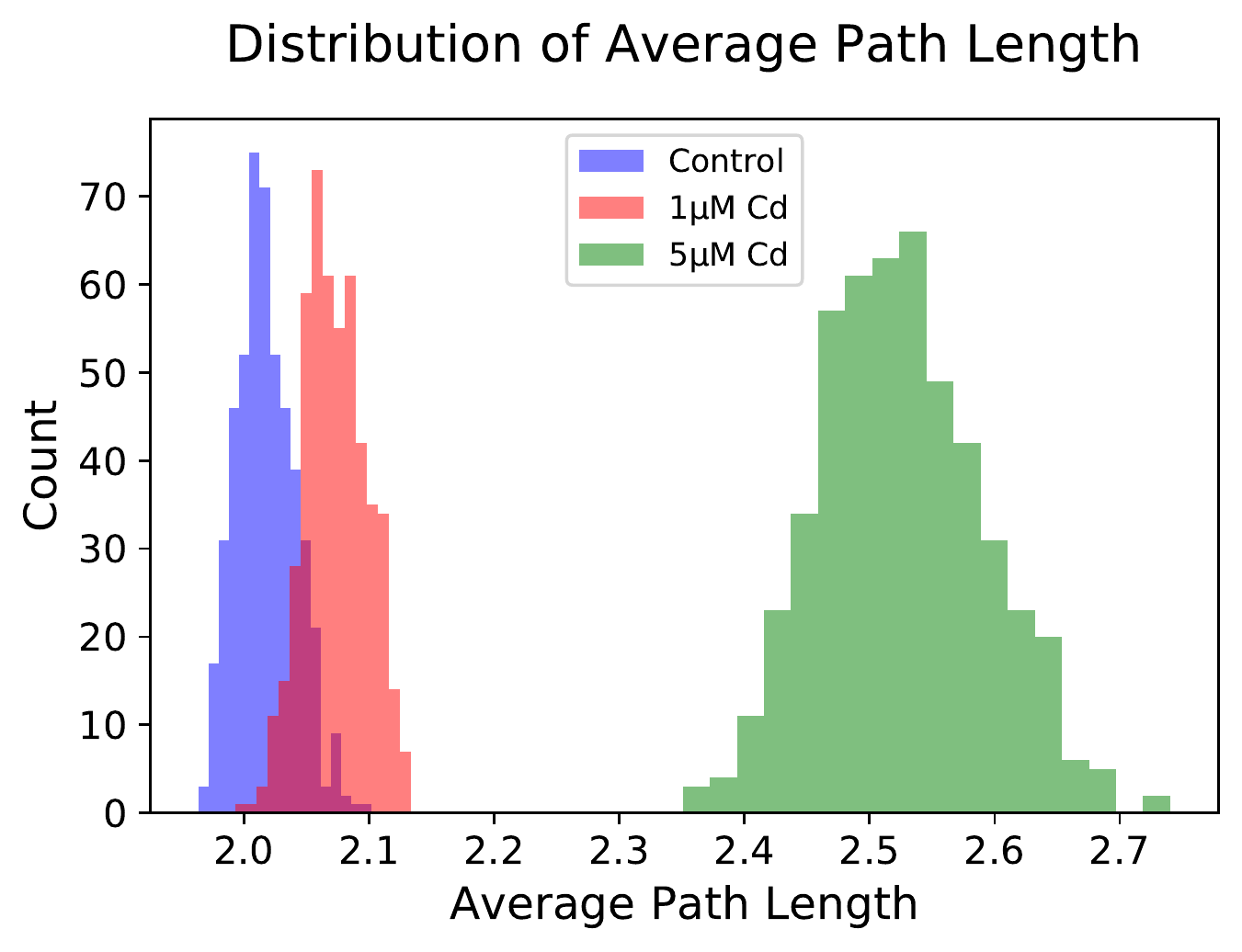}
  \caption{Posterior of average path length distribution with Cadmium concentration. The average path length increases significantly with the increase of cadmium concentration.}
  \Description{Number of Connections Distribution with Cadmium Concentration}
  \label{fig:Distribution_APL}
\end{figure}

\begin{figure}[ht]
  \centering
  \subfloat{%
    \includegraphics[width=0.93\linewidth]{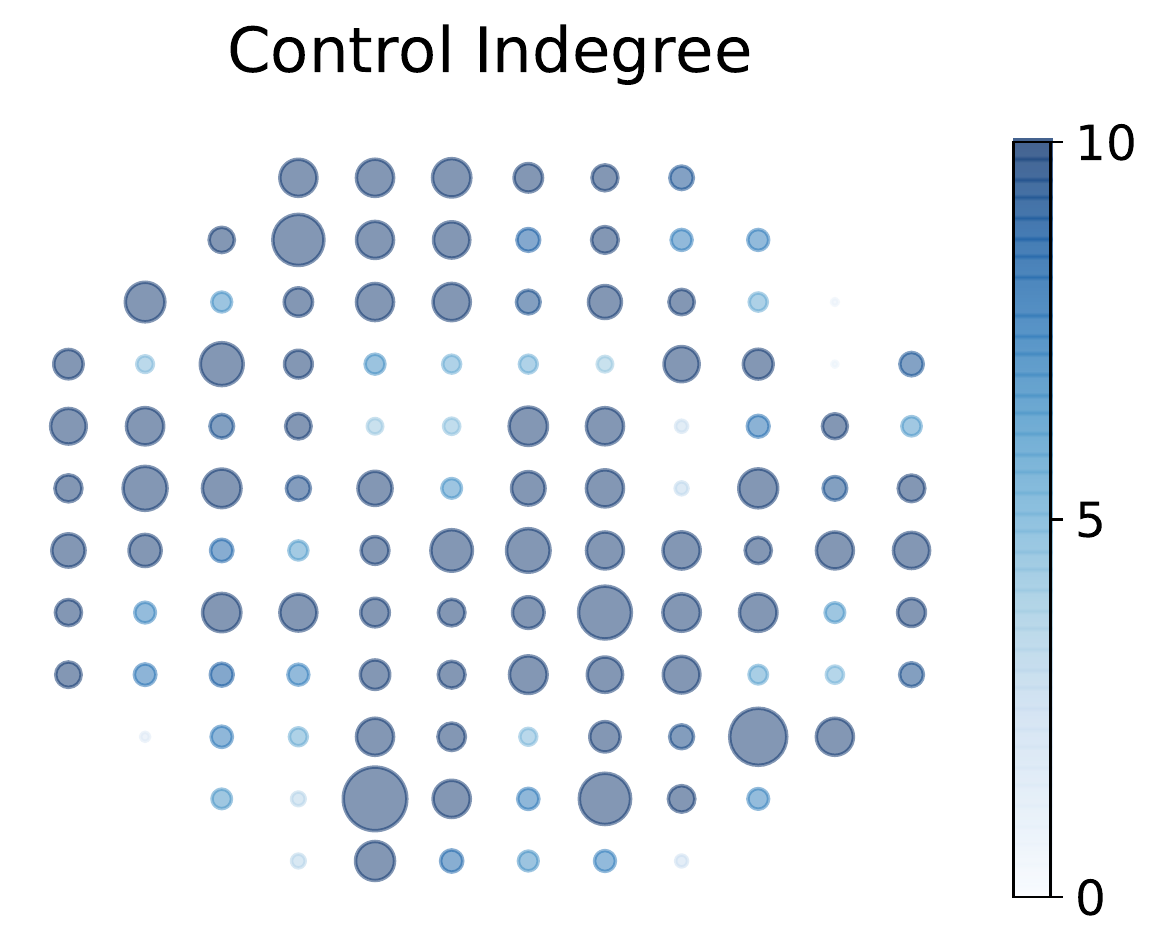}%
  } \quad
  \subfloat{%
    \includegraphics[width=0.93\linewidth]{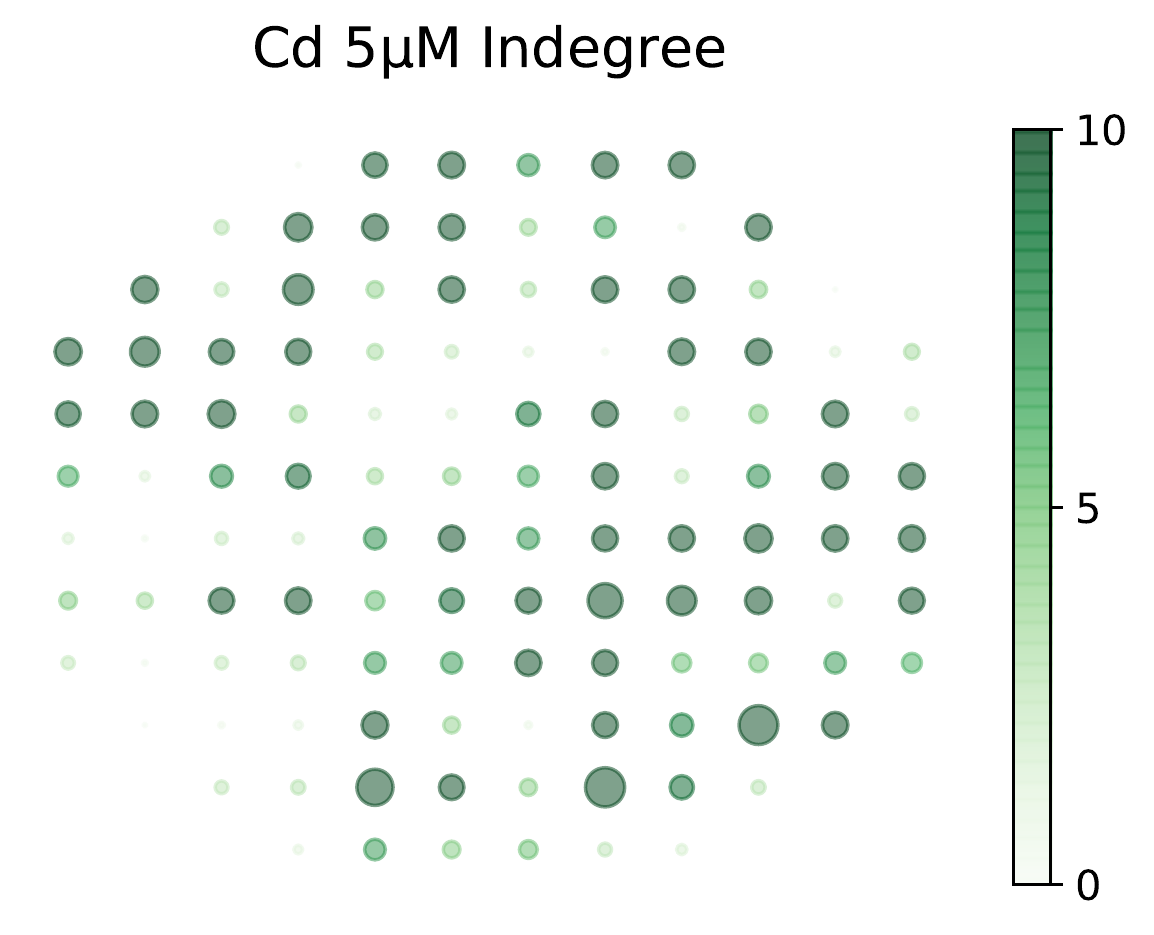}%
  }
  \caption{Incoming connections for control and Cadmium 5$\mu M$. Nodes are sized and colored according to the number of incoming connections. The overall connection pattern remains but the indegree decreases prominently with the introduction of Cadmium into the culture.}
  \Description{Incoming connections for control(a) and Cadmium 5uM(b). Nodes are sized and colored (online) according to the number of incoming connections. The overall connection pattern remains but the indegree decreases prominently with the introduction of Cadmium into the culture.}
  \label{fig:Scatter_indegree}
\end{figure}

\begin{figure}[ht]
  \centering
  \subfloat{%
    \includegraphics[width=0.93\linewidth]{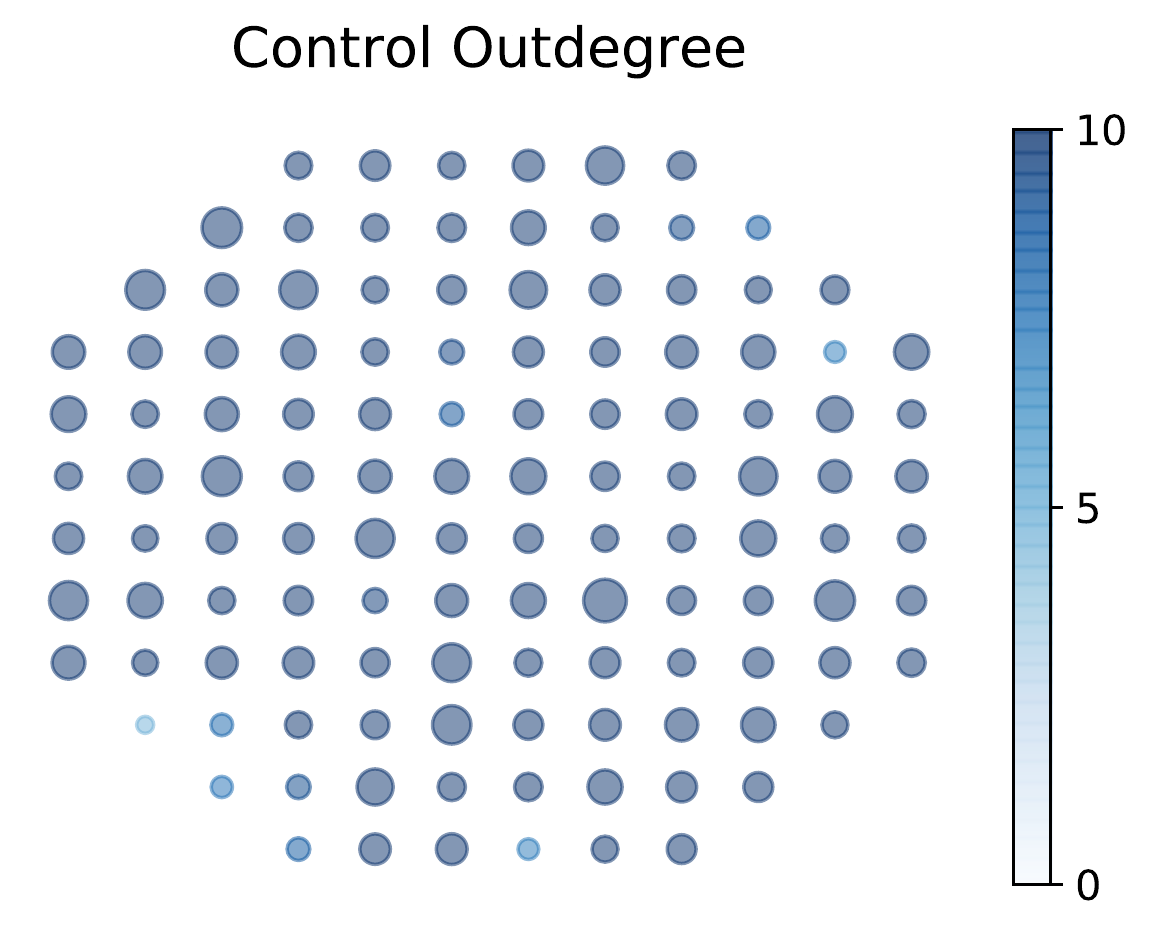}%
  } \quad
  \subfloat{%
    \includegraphics[width=0.93\linewidth]{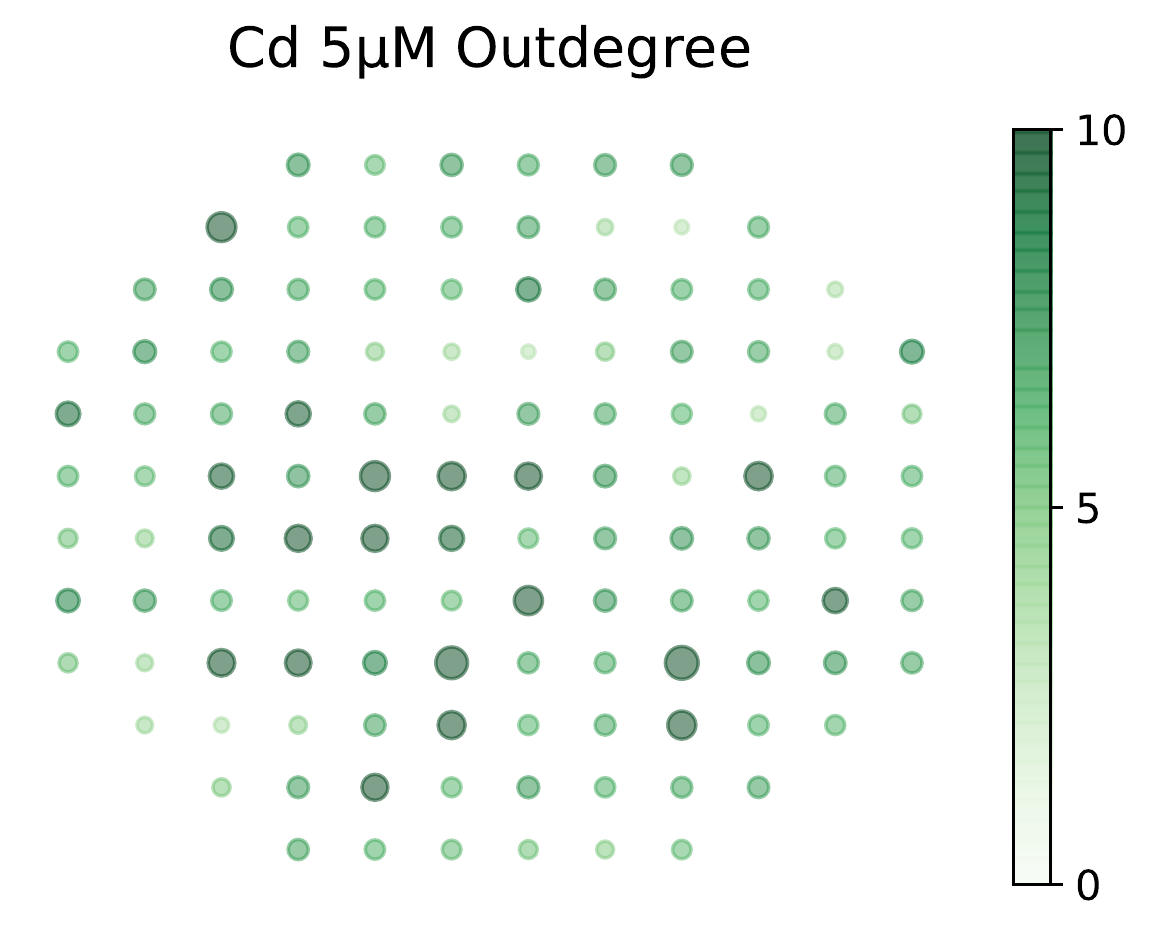}%
  }
  
  \caption{Outgoing connections for control and Cadmium 5$\mu M$. Nodes are sized and colored according to the number of outgoing connections. The overall connection pattern remains but the outdegree decreases prominently with the introduction of Cadmium into the culture.}
  \Description{Incoming connections for control(a) and Cadmium 5uM(b). Nodes are sized and colored (online) according to the number of incoming connections. The overall connection pattern remains but the indegree decreases prominently with the introduction of Cadmium into the culture.}
  \label{fig:Scatter_outdegree}
  
\end{figure}

Cadmium is a toxic heavy metal that accumulates in living systems and is currently one of the most important occupational and environmental pollutants~\cite{branca2018cadmium}. Cells have a calcium signalling toolkit whose components can be mixed and matched to create various spatial and temporal signals~\cite{berridge2000versatility}. Cadmium can change the intracellular concentration of calcium, which is a universal and versatile intracellular messenger~\cite{choong2014interplay}.

We reduced the release probability of presynaptic vesicles by titrating in increasing amounts of cadmium in order to decrease the influx of calcium current into the presynaptic terminal. Baseline recordings were performed 5 minutes prior to any cadmium addition.  Following the baseline recording, the cadmium concentration was brought to 1$\mu M$ and the solution was mixed gently. The MEA was placed on the recording headstage and allowed to equilibrate for 3 minutes, followed by a 3-minute recording.  The same procedure was performed for 5$\mu M$ cadmium in the same MEA.

To analyze how cadmium affects the structure of neuronal networks, we compared the obtained posterior matrices $A$ and $W$ for a set of MEA recordings, comprised of one control with no cadmium introduced and 2 others with 1$\mu M$ and 5$\mu M$ of cadmium. We used a threshold of 0.05 (the value is chosen based on the outlier of the weight distribution) to eliminate those tiny weights in the inferred weight matrix, which is noise in biology.
To understand the topology, we look at the following graph metrics for each posterior sample: number of connections, average clustering coefficient, and average path length. Overall, we find that Cadmium does change the topology of the estimated networks with statistical significance in posterior regions. As is shown in Fig.~\ref{fig:Distribution}, the mean connection counts decrease with introduction of cadmium to the culture. Higher concentrations of cadmium lead to a drastic decrease in the number of connections in the functional networks.

Clustering coefficient is the number of edges between a electrode’s immediate neighbors divided by all possible connections that could exist among them. It measures the level of local connectivity between electrodes. In Fig.~\ref{fig:Distribution_CC}, average clustering coefficient decreases with statistical significance when cadmium is in the culture, indicating that cadmium cultures are less likely to have connections within tightly connected groups or communities.

Average path length is the average shortest path length for all possible pairs of electrodes. Fig.~\ref{fig:Distribution_APL} shows that the average path length increases significantly with the increase of cadmium concentration. That validates the idea that cadmium can impede neuron signal transmissions by changing the intracellular concentration of calcium.

In Fig.~\ref{fig:Scatter_indegree}, we compare the indegree of each electrode for control and 5$\mu M$ Cadmium concentration. We can see that the overall connection pattern maintains but the indegree decreases prominently with the introduction of Cadmium into the culture. Similar phenomenon can be detected for outdegree of each electrode for control and 5$\mu M$ Cadmium concentration in Fig.~\ref{fig:Scatter_outdegree}.

\section{related work}\label{related}
MEAs provide opportunities for researchers to understand neuronal connectivity, by recording spike trains but also presents severe data analysis challenges. Inferring functional connectivity networks is critical to many applications in neuroscience. To analyze high-dimensional spike trains, there exist several methods to measure connection weights. For example, CCG~\cite{garofalo2009evaluation,salinas2001correlated}
is a metric that was built from two electrodes’ spike trains. It presents the probability of an electrode firing to a spike in a $\tau$ milliseconds time window before or after another electrode fires. MIC has been widely used to identify known and novel relationships for data sets in gene expression and global health~\cite{MIC}. MIC is a two-variable dependent measure that captures functional relationship, by  providing a score that roughly equals the coefficient of determination of the data relative to the regression function. Biophysically-inspired metrics extracts a directed functional connectivity matrix, based on the spike train~\cite{bio-inspired}. It uses exponential decay property in axon potential signals to quantify the connection coefficients. However, despite the popularity of applying those metrics on MEA recordings, we can not rely on them to get reliable functional networks because they are deterministic and there is no model for the data.

GLM, a commonly used modeling framework~\cite{chen2013overview, pillow2008spatio}, can model the binary fire/not (1/0) and spike train recordings. We can use the $logit$ link function to model the binomial distribution. GLMs are often fit by maximum a posteriori (MAP) estimation~\cite{paninski2004maximum, truccolo2005point}. However, when it comes to high-density recordings, MAP cannot fully convey the information in the posterior with a point estimate. Bayesian inference is a process using Bayes' theorem based on the prior beliefs about the probabilistic data generation process~\cite{parr2018computational}.  It updates the posterior for parameters in data generation model, as more data becomes available. GLM and graph-based models have been combined to infer the neuron connectivity patterns in a Bayesian way~\cite{Linderman}. However, the Bayesian techniques have a considerable downside of the increased computation time, especially when the number of electrodes is large. In this paper, we introduced a scalable Bayesian inference framework on large scale MEA recordings. The functional connectivities are inferred from a joint probabilistic model of GLM and networking. 


\section{Discussion}\label{discussion}
In this paper we have focused on inferring functional neuronal network connections using MEA recordings. Specifically, our goal has been to infer the functional connectivity for MEAs with large numbers of probes. Along this line, we proposed a scalable Bayesian inference framework. Our framework makes use of the hierarchical structure of networks of neurons and splits the whole array into smaller local networks for network inference. The splitting strategy decreases the average sampling time quadratically with the number of sub-regions.
We also provide a strategy for inferring the connectivity between local networks. By comparing with ground truth for both synthetic data and real world human expert labeled MEA recordings, our experimental results provide compelling evidence that our framework can infer the underlying functional connectivity.  Furthermore, we applied the proposed framework to a controlled cadmium dataset, and the results confirm its utility. As the density of the MEA continues to increase, our method will become more valuable to be able to infer the neuronal network structure efficiently.

Our experimental results demonstrate the usefulness of this framework. Here we suggest some avenues for future work. First, the model could be extended to account for more detailed biological knowledge.  For example, there are different types of neurons such as motor neurons and interneurons, which exhibit different types of behaviors in response to incoming signals.  The auto-regressive propensity model, which is currently linear, can also be improved to incorporate nonlinear effects or other mechanistic firing models. Secondly, developing a method to determine when and how to split can allow for the framework to be used more robustly. Moreover, using the framework to study different factors instead of Cadmium, such as the presence of certain genes, may produce a greater understanding of the biologically complex systems.
 


\bibliographystyle{ACM-Reference-Format}
\bibliography{ref}


\newpage
\begin{appendices}
\section{appendix for reproducibility}
To support the reproducibility of the results in this paper, we provide the details in our experiments. We use 24 CPUs with model of Intel(R) Xeon(R) CPU E5-2670 v3 @ 2.30GHz. We perform parallel sampling with OpenMP and the average sampling time scales with the number of CPUs~\cite{Linderman}.

The network prior does affect the inference of $A$ and $W$ but in an indirect way, which is small compared to the effect of the data. For all the results in this paper, we ran the Gibbs sampler for 1000 iterations and verified that the results stayed consistent. 
Hyper-parameters we used for synthetic and real data in Tab.~\ref{tab:append1} and Tab.~\ref{tab:append2} respectively.

\subsection{Hyper-parameters for synthetic experiments}

The hyper-parameters in Tab.~\ref{tab:append1} are fixed for synthetic experimental results in Fig.~\ref{fig:Synthetic4}, Fig.~\ref{fig:Synthetic10},  Fig.~\ref{fig:Synthetic2nd} and Tab.~\ref{tab:Non_overlap} and Tab.~\ref{tab:overlap}.

\subsection{Hyper-parameters for real data experiments}
The hyper-parameters we used for inference on real MEA recordings in Fig.~\ref{fig:Compare}, Fig.~\ref{fig:Distribution}, Fig.~\ref{fig:Distribution_CC}, Fig.~\ref{fig:Distribution_APL}, Fig.~\ref{fig:Scatter_indegree} and Fig.~\ref{fig:Scatter_outdegree} are listed in Tab.~\ref{tab:append2}. We chose the hyper-parameters based on the choice in ~\cite{Linderman} and the observed firing probability in biological experiments.

\begin{table}
  \caption{Hyper-parameters for synthetic experiments}
  \label{tab:append1}
  \begin{tabular}{cccl}
    \toprule
    T&$\mu_{w_n}$&$S_{b_n}$\\
    \midrule
    100&1&1\\
    									
  \bottomrule
\end{tabular}
\end{table}

\begin{table}
  \caption{Hyper-parameters for real data experiments}
  \label{tab:append2}
  \begin{tabular}{ccccccl}
    \toprule
    T&$\mu_{w_n}$&$S_{b_n}$&$S_{w_n}$&$\mu_b$&$\rho$\\
    \midrule
    100&1&1&1&-2&0.1\\
    									
  \bottomrule
\end{tabular}
\end{table}

\end{appendices}

\end{document}